\UseRawInputEncoding
\documentclass[prl,twocolumn,superscriptaddress,floatfix,nopacs,longbibliography]{revtex4-2}
\usepackage{graphicx,amsfonts,amssymb,amsmath,hyperref,enumerate,footnote}
\usepackage{graphicx}
\usepackage{float}
\usepackage{indentfirst}
\usepackage{verbatim}
\usepackage{tikz}
\usepackage{ragged2e}
\usetikzlibrary{quantikz}
\usetikzlibrary{positioning}
\usepackage{array}
\usepackage[utf8]{inputenc}
\usepackage{mathtools}
\usepackage{blindtext}
\usepackage{xcolor}
\usepackage{tikz}
\usepackage{tabularx}
\usepackage{multirow}

\usetikzlibrary{arrows.meta,
                chains,
                positioning,
                shapes.geometric}
\usepackage{siunitx}
\tikzstyle{startstop} = [rectangle, rounded corners, minimum width=3cm, minimum height=1cm,text centered, draw=black, fill=red!30]
\tikzstyle{io} = [trapezium, trapezium left angle=70, trapezium right angle=110, minimum width=3cm, minimum height=1cm, text centered, draw=black, fill=blue!30]
\tikzstyle{process} = [rectangle, minimum width=3cm, minimum height=1cm, text centered, draw=black, fill=orange!30]
\tikzstyle{decision} = [diamond, minimum width=3cm, minimum height=1cm, text centered, draw=black, fill=green!30]
\tikzstyle{arrow} = [thick,->,>=stealth]

\newif\ifhyper
\hypertrue
\ifhyper
\hypersetup{
   citecolor = {green},
   colorlinks = {true}, % false
   urlcolor = {blue} % magenta
}
\fi

\newcommand{\beq}{\begin{equation}}
\newcommand{\eeq}{\end{equation}}
\newcommand{\beqa}{\begin{eqnarray}}
\newcommand{\eeqa}{\end{eqnarray}}

\def\ket#1{\vert#1\rangle}

\def\Longarrow{\protect\@lra}
\def\@lra{\relbar\joinrel\relbar\joinrel\relbar\joinrel%
          \relbar\joinrel\rightarrow}

\begin{document}
\title{Projected Entangled Pair States with flexible geometry}

\author{Siddhartha Patra}
\affiliation{Donostia International Physics Center, Paseo Manuel de Lardizabal 4, E-20018 San Sebasti\'an, Spain}
\affiliation{Multiverse Computing, Paseo de Miram\'on 170, E-20014 San Sebasti\'an, Spain}

\author{Sukhbinder Singh}
\affiliation{Multiverse Computing, Spadina Ave., Toronto, ON M5T 2C2, Canada}

\author{Rom\'an Or\'us}
\affiliation{Donostia International Physics Center, Paseo Manuel de Lardizabal 4, E-20018 San Sebasti\'an, Spain}
\affiliation{Multiverse Computing, Paseo de Miram\'on 170, E-20014 San Sebasti\'an, Spain}
\affiliation{Ikerbasque Foundation for Science, Maria Diaz de Haro 3, E-48013 Bilbao, Spain}

\begin{abstract}
Projected Entangled Pair States (PEPS) are a class of quantum many-body states that generalize Matrix Product States for one-dimensional systems to higher dimensions. In recent years, PEPS have advanced understanding of strongly correlated systems, especially in two dimensions, e.g., quantum spin liquids. Typically described by tensor networks on regular lattices (e.g., square, cubic), PEPS have also been adapted for irregular graphs, however, the computational cost becomes prohibitive for dense graphs with large vertex degrees. In this paper, we present a PEPS algorithm to simulate low-energy states and dynamics defined on arbitrary, fluctuating, and densely connected graphs. We introduce a cut-off, $\kappa \in \mathbb{N}$, to constrain the vertex degree of the PEPS to a set but tunable value, which is enforced in the optimization by applying a simple edge-deletion rule, allowing the geometry of the PEPS to change and adapt dynamically to the system's correlation structure. We benchmark our flexible PEPS algorithm with simulations of classical spin glasses and quantum annealing on densely connected graphs with hundreds of spins, and also study the impact of tuning $\kappa$ when simulating a uniform quantum spin model on a regular (square) lattice. Our work opens the way to apply tensor network algorithms to arbitrary, even fluctuating, background geometries.
\end{abstract}

\maketitle

\textit{Introduction.--} 
In recent decades, tensor network algorithms have become state-of-the-art numerical methods for simulating complex many-body systems \cite{PEPSReviewOrus, PEPSReviewCirac}. The most famous and successful example of a tensor network is the Matrix Product State (MPS) \cite{Fannes}, which has been established as a benchmark or simulating one-dimensional (1D) quantum systems. The Projected Entangled Pair States (PEPS) method \cite{PEPS} extends the MPS framework to effectively simulate two-dimensional (2D) and higher-dimensional quantum systems, possibly with an arbitrary geometry \cite{gPEPS}, but incurs a significantly higher computational cost. Despite the limiting cost, PEPS have significantly advanced our understanding of strongly correlated quantum systems, in particular those with topological order \cite{PEPSClassifyingPhases, PEPSTopologicalOrder}, including quantum spin liquids \cite{PEPSSpinLiquid}.

\begin{figure}
    \centering
    \includegraphics[width=\columnwidth]{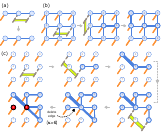}
    \caption{(a) An MPS updated after absorbing a two-site gate. (b) Two-site gates applied on the standard, fixed PEPS update the bonds but not the geometry of the tensor network. (c) A sequence of two-site gates is applied on the flexible PEPS, when evolving under an irregular densely connected Hamiltonian. Each update adds an edge to the PEPS. After a few updates, a tensor (no. 5) acquires $5 > \kappa = 4$ edges, triggering the deletion of the indicated edge and updating tensors 4 and 5. The bond to be deleted is selected according to bond entanglement entropy score ( = edge thickness).}
    \label{fig:main_fig1}
\end{figure}

PEPS are usually applied to approximate low-energy states or dynamics of Hamiltonians that are the sum of local terms, namely, interaction terms that involve only a few neighbouring sites. One constructs the PEPS tensor network according to a graph $\mathcal{P}$ (the ``correlation geometry'') that best reflects the (assumed) correlation structure of the target state. When the Hamiltonian is the sum of only two-site interactions, such as the Ising model, one can consider the Hamiltonian geometry $\mathcal{G}$, namely, the graph whose nodes and edges correspond to the degrees of freedom and the two-site interaction terms in the Hamiltonian, respectively. Typically, $\mathcal{P}$ is chosen to match the Hamiltonian geometry $\mathcal{G}$. In this case, when $\mathcal{P} = \mathcal{G}$, the computational cost of the PEPS roughly scales as $\approx \chi^K$, where $\chi$ is the PEPS bond dimension (namely, the maximum size of any bond index of the PEPS) and $K$ is the maximum degree of any vertex in the Hamiltonian graph $\mathcal{G}$, which quickly becomes prohibitive when the Hamiltonian is densely connected, i.e., for large $K$. The choice $\mathcal{P} = \mathcal{G}$ seems reasonable for translation-invariant Hamiltonians defined on regular lattices, in which case all the terms of the Hamiltonian are comparable. However, for Hamiltonians with disorder or those defined on irregular graphs --- for example, classical and quantum spin glasses \cite{SpinGlass1, SpinGlass2, SpinGlassBook}, which are relevant also for combinatorial optimization problems \cite{SpinGlassCombinatorialOptimization}, Sachdev-Ye-Kitaev models \cite{SYK1, SYK2}, and quantum chemistry models (see, e.g. \cite{QuantumChemistryReview, QuantumChemistry}) --- the correlation geometry of the target state could significantly deviate from the Hamiltonian geometry. Even for certain translation-invariant models on regular lattices in higher dimensions ($d \ge 4)$ where mean-field techniques become more effective, or models on locally tree-like lattices \cite{ibm_simulator}, one might expect that a more sparsely connected $\mathcal{P}$, reflecting small correlations, should suffice. However, guessing an effective PEPS geometry in such situations is a difficult task, and presents a key challenge for PEPS methods based on a fixed geometry. Additionally, we know of many situations where there is no underlying lattice at all, e.g., when simulating a quantum circuit, or for the case of a dynamical system with fluctuating background geometry.  

While attaching a PEPS to a predefined lattice can be useful, this is also a stringent constraint. In general, it would be more beneficial to let the geometry fluctuate and adapt dynamically, as the algorithm flows, to the entanglement structure that is natural to the system. 

In this paper we materialize the above intuition by proposing a PEPS algorithm based on a \textit{flexible} geometry for simulating models defined on irregular and densely connected graphs with large $K$, even comparable to the system size. In order to control the exponentially scaling cost $\approx \chi^K$, we introduce a new PEPS cutoff $\kappa \in \mathbb{N} \ll K$. At any step in the optimization, we enforce that the maximum degree of the PEPS tensors (= max. vertex degree of $\mathcal{P}$) remains $\le \kappa$, but otherwise allow the PEPS geometry change and adapt to the (generally unknown) target correlation geometry. To this end, we incorporate a simple scheme to \textit{delete} edges (tensor indices) during the PEPS optimization. Starting, for instance, with a product state, a PEPS variational optimization or time-evolution proceeds by adding new edges (and correlations) to the tensor network. Whenever a tensor violates the $\kappa$-cutoff, we delete a bond index from that tensor, chosen according to a bond entanglement score. We propose a simple edge-deletion rule, incorporated with PEPS optimization based on \textit{simple updates} \cite{SimpleUpdate}, and demonstrate that it gives remarkably accurate results across several benchmarking simulations of irregular and densely connected models with hundreds of spins, up to approximately 1200 spins. Nonetheless, our edge deletion rule and its generalizations can also be implemented along with systematic improvements of the PEPS optimization by using cluster and full-updates \cite{PEPSUpdates1, PEPSUpdates2}, see Appendix \ref{app:pepsalgorithms}. % The Appendices briefly 

\textit{Setup.--} 
Consider a many-body system made of $L$ sites, described by a tensor product Hilbert space  $\mathbb{V} \cong \bigotimes_{i=1}^{L} \mathbb{V}_i$ and a local Hamiltonian $H:\mathbb{V}\rightarrow\mathbb{V}$,
which we assume is the sum of only two-site terms, $H = \sum_{ij} h_{ij}$, where $h_{ij}$ is a hermitian operator $ \mathbb{V}_i \otimes \mathbb{V}_j \rightarrow \mathbb{V}_i \otimes \mathbb{V}_j$ on sites $i, j$ times the Identity on remaining sites. The total Hamiltonian $H$ defines a graph $\mathcal{G}$ whose nodes correspond to lattice sites and sites $i,j$ are connected with an edge only if an interaction term $h_{ij}$ appears in the Hamiltonian.

\textit{Aims.--}
Our goal here is to approximate the low-energy states and real-time dynamics of the system using PEPS-based time evolution methods \footnote{For concreteness, we focus on PEPS time-evolution algorithms to approximate the low-energy states and dynamics of the system, but our flexible PEPS updates can also be incorporated in PEPS-based variational algorithms}. Specifically, we would like to approximate the state $\ket{\Psi} \equiv e^{-\tau H} \ket{\Psi_0}$ (up to normalization), where the initial state $\ket{\Psi_0}$ is a product state or a simple PEPS with a small bond dimension. Here, $\tau$ is either real, in which case $\ket{\Psi}$ is a low-energy state (e.g., the ground state) resulting from an imaginary time evolution, or purely imaginary, $\tau = iT$, which corresponds to a real-time evolution for time $T$. Using Suzuki-Trotter decomposition, we approximate the time evolution operator $e^{-\tau H}$ as a circuit comprised of two-site gates, $e^{-\tau H} \approx \Pi_{ij} g_{ij}$, where the gates $g_{ij} \equiv e^{-\delta \tau h_{ij}}$ are connected according to graph $\mathcal{G}$ and sliced into $N$ time steps $\delta \tau = N / \tau$. At each trotter-step, the PEPS is updated after absorbing a time-slice of gates. The Suzuki-Trotter decomposition is \textit{exact} if all the two-site terms commute but otherwise incurs an error that scales polynomially with $\delta \tau$. In order to minimize the Suzuki-Trotter error, we partition the circuit into $G$ ``layers'', $\{\mathcal{G}_k\}_{k=1}^{G}$, such that each layer $\mathcal{G}_k$ contains only non-overlapping edges\footnote{Consider a graph with 6 nodes numbered 0,1,...,5, and edges: [(0,1), (1,2), (2,3), (3,4), (4,5), (0,3), (2,4), (3,5)]. We can group these edges into four layers of non-intersecting edges:
Layer1=[(0,1), (2,3), (4,5)]
Layer2=[(1,2), (3,4)]
Layer3=[(0,3), (2,4)]
Layer4=[(3,5)]. Note that for several problems studied in this paper we also tested the straightforward gate sequence $(0,1),(0,2)...(1,2),(1,3)...$ with with a small trotter step $= 0.01$. Surprisingly, we obtained comparable results despite incurring an uncontrolled trotter error.}, so that all the two-site gates within any layer commute with one another. Thus, we obtain the circuit:
%\begin{equation}\label{eq:trotter}
$e^{-\delta \tau H} \approx \prod_{k = 1}^G \left(\bigotimes_{(i,j) \in \mathcal{G}_k} g_{ij}\right)$.
%\end{equation}

\textit{Flexible PEPS algorithm.--} We now describe our new PEPS algorithm to approximate the trotterized time evolution from state $\ket{\Psi_0}$ towards $\ket{\Psi}$. At any step in the evolution, the instantaneous state is approximated as a PEPS $\ket{\mathcal{P}_t} \in \mathbb{V}$ whose geometry $\mathcal{P}_t$ is now allowed to change along the time-evolution such that $\ket{\Psi_0} = \ket{\mathcal{P}_{0}}$ and $\ket{\Psi} \approx \ket{\mathcal{P}_{N}}$. We choose $\ket{\Psi_0} = \ket{\mathcal{P}_0}$ to be a product state, corresponding to an edge-less graph $\mathcal{P}_0$. The initial evolution proceeds by adding bonds to the PEPS. Specifically, gate $g_{ij}$ creates entanglement between sites $i$ and $j$, captured by creating a bond between tensors located at those sites. The PEPS at time $t$ consists of a set of tensors $\{\Gamma^{[i]}_t\}_{i=1}^{L}$ and diagonal matrices $\{\lambda^{[i,j]}_t\}_{i,j}$, attached to the nodes $i$ and edges $[i,j]$ of the graph $\mathcal{P}_t$, respectively, and represents a state $\ket{\mathcal{P}_t} \in \mathbb{V}$, which can be recovered by contracting all the PEPS tensors. The basic optimization step consists of updating the PEPS tensor network after applying a single two-site gate. In this paper, we considered the well-known \textit{simple update} scheme, see Appendix \ref{app:pepsalgorithms}, absorbing the gate locally into the PEPS and ignoring the rest of the system, which already gave accurate results across a range of benchmarking simulations. As the gates are applied, the PEPS becomes increasingly more connected, thus, escalating the computational cost of the updates, rapidly becoming prohibitive for densely connected Hamiltonians. In order to control this cost, we now enforce a vertex constraint, namely, no PEPS tensor is allowed to have more than $\kappa \in \mathbb{N}$ bond indices at any time in the evolution. If after applying a gate a tensor violates this constraint, we proceed to remove a bond index from that tensor. In order to reduce the impact on expectation values of local observables acting at that site, we propose to remove the bond with minimal \textit{bond entanglement entropy} (BEE) $\mathcal{E}_{i,j} \equiv -\sum_m \lambda^{[i,j]}_{m} \log_2(\lambda^{[i,j]}_{m})$, where $\lambda^{[i,j]}_{m}$ are the entries of the diagonal bond matrix $\lambda^{[i,j]}$ \footnote{We also tried a slightly different edge-deletion strategy. If the new bond resulting from the current gate update has bond entanglement in the lowest $30\%$ then we remove that bond. Otherwise, we remove the bond with the least bond entanglement entropy. This strategy worked better for some models.}. In order to delete index $a$ between nodes $x$ and $y$, we truncate the bond matrix $\lambda^{[x,y]}$ keeping only the \textit{largest} singular value and the corresponding left and right singular vectors, which are then absorbed into the vertex tensors $\Gamma^{[x]}, \Gamma^{[y]}$  as
\begin{align}
\Gamma^{[x]}_{\ldots,~a,~\ldots} &\rightarrow (\lambda^{[x,y]}_{\mbox{\tiny max}})^{1/2}~\Gamma^{[x]}_{\ldots,~a=1,~\ldots},\label{eq:delete1}\\
\Gamma^{[y]}_{\ldots,~a,~\ldots} &\rightarrow (\lambda^{[x,y]}_{\mbox{\tiny max}})^{1/2}~\Gamma^{[y]}_{\ldots,~a=1,~\ldots}.\label{eq:delete2}
\end{align}
(We assume here that values of index $a$ correspond to singular values sorted in descending order, i.e., $a=1$ corresponds to the largest value.) Slicing and scaling tensors, Eqs.~\ref{eq:delete1}-\ref{eq:delete2}, incurs minimal computational cost. Therefore, the cost of the flexible PEPS algorithm is still dominated by the cost of the PEPS updates, roughly scaling as $\approx \chi^\kappa$ (compared to $\approx \chi^K$ in a fixed PEPS setting).

Though heuristic, we found that the BEE is a meaningful edge-deletion score. First, we demonstrate later that our edge-deletion rule has the intuitively expected impact when applied to uniform models on \textit{regular} lattices, in particular, BEE can be used to locate critical points in such models with reasonable accuracy. Second, for several densely connected Hamiltonians, we observed that the ordered \textit{bond spectrum} $S_i \equiv \{\mathcal{E}_{i,j}\}_j$ of site $i$ -- namely, BEE values of all the bonds at $i$ -- decays exponentially with the vertex degree, thus, the impact of the BEE-based truncation is potentially controllable.

\textit{Results.---}
We now demonstrate that our flexible PEPS algorithm based on simple updates + simple measurements (see Appendix \ref{app:pepsalgorithms}), works reasonably well for simulations of (1) classical Ising spin glasses, (2) disordered quantum annealing, and (3) the uniform 2D quantum Ising model.

\begin{figure}
    \centering
    \includegraphics[width=\columnwidth]{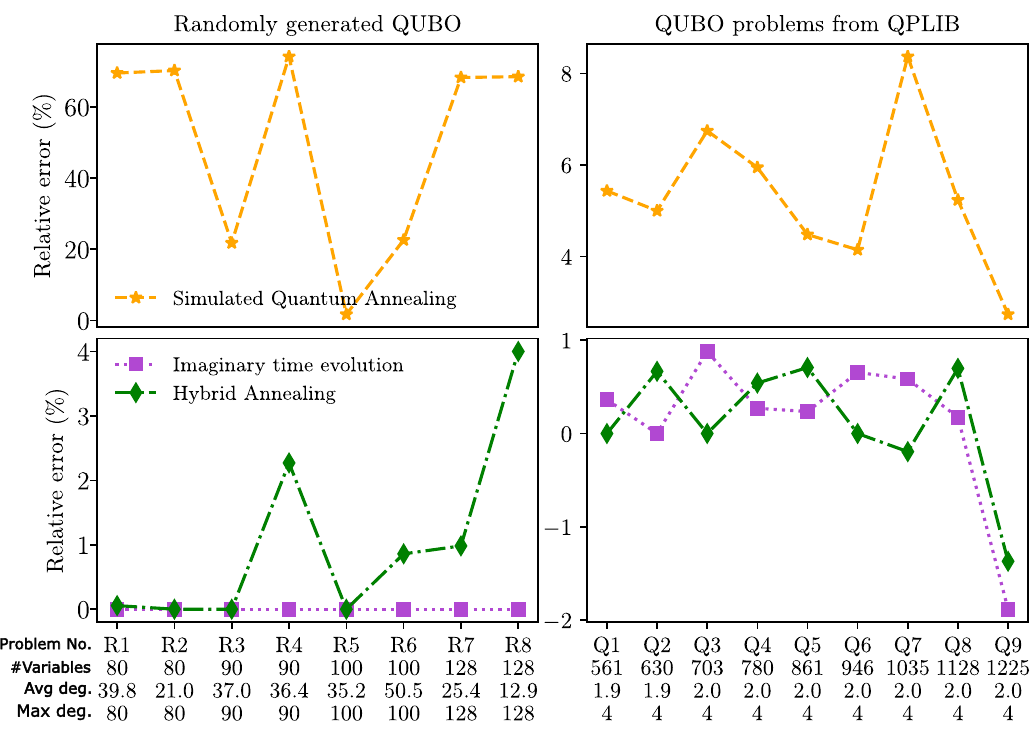}
    \caption{Relative error ($\%$) of the flexible PEPS algorithm vs. Gurobi for QUBO problems (details listed along $x$-axis) when using imaginary time-evolution, quantum annealing, and hybrid annealing. Simulation details in Appendix \ref{app:simdetails}.}
    \label{fig:main_results}
\end{figure}

\textit{(1) Classical Spin Glasses.} We applied the flexible PEPS algorithm to find the ground of a classical Ising spin glass defined on a graph $\mathcal{G}$,
$H_{\mbox{\tiny  Ising-SG}} \equiv \sum_i h_i \sigma^{i}_z + \sum_{ij} J_{ij} \sigma^{i}_z \sigma^{j}_z + C, \label{eq:spinglass}
$, where $i,j$ enumerate the vertices of $\mathcal{G}$, each vertex corresponds to a site described by $\mathbb{C}^2$, $\sigma_z = \big(\begin{smallmatrix}
  1 & 0\\
  0 & -1
\end{smallmatrix}\big)$ is the Pauli $Z$ matrix, and $C$ is a constant. All the eigenstates of $H_{\mbox{\tiny  Ising-SG}}$ are product states, since $H_{\mbox{\tiny  Ising-SG}}$ is already diagonal in the $Z$ basis on each site. Finding the ground product state of this model is equivalent to solving a generally NP-hard \cite{SpinGlassNPHard}  Quadratic Unconstrained Binary Optimization (QUBO) problem, corresponding to minimizing the cost function $F = \sum_{ij} x_iQ_{ij}x_j$, where $x_i\in [0,1]$ is a (classical) binary variable, and $Q$ is the quadratic cost matrix. The classical Ising spin glass $H_{\mbox{\tiny  Ising-SG}}$ maps to a QUBO by identifying $\sigma_i = (x_i-1/2)$, $h_i=(Q_{ii}+\sum_j Q_{ij}/2)$, $J_{ij}=Q_{ij}$, and $C=\sum_{ij}(Q_{ii}/2+Q_{ij}/4)$. The binary string $x_0, x_1, \ldots, x_L$ that minimizes $F$ corresponds to the product state $\ket{x_0} \otimes \ket{x_1} \ldots \otimes \ket{x_L}$ with minimum energy. We approximated the ground state $|\psi_G\rangle$ of $H_{\mbox{\tiny  Ising-SG}}$ by simulating imaginary-time evolution,
$|\psi_G\rangle = \lim_{N\rightarrow \infty} \frac{e^{-N \delta \tau \beta H }|+\rangle}{|| e^{- N \delta \tau \beta H }|+\rangle ||}$, using the flexible PEPS; here, $\ket{+} = \bigotimes_i \left(\frac{1}{\sqrt{2}}\ket{0}_i + \frac{1}{\sqrt{2}}\ket{1}_i\right)$. The evolution operator was trotterized as explained previously, and the PEPS was optimized by combining simple-updates with the BEE edge-deletion rule. The algorithm often converged to an entangled PEPS with a small bond dimension (= 2 or 3), which we expected to have a large overlap with the ground product state. We sampled 100 product states from this entangled PEPS -- adapting the perfect sampling algorithm for the MPS \cite{MPSSampling1, MPSSampling2} to our flexible PEPS setting, see Appendix \ref{app:sample} -- and chose the product state with the lowest energy as the solution.

We approximated the ground state of $H_{\mbox{\tiny  Ising-SG}}$ for several values of the couplings $h_i$ and $J_{ij}$. We chose eight randomly sampled couplings, corresponding to models R1, R2, $\ldots$, R8, ranging between 80 and 128 spins, and nine problems from the QPLIB database \cite{QPLIB} corresponding to the models Q1, Q2, $\ldots$, Q9, ranging between 561 and 1225 spins, see Appendix \ref{app:problemset} for details. QPLIB problems have a low vertex degree ($\le 4$) but are known to be hard problems. Fig.\ref{fig:main_results} (bottom) and Table \ref{tab:error_values} show the error of the ground state energy relative to the solution obtained using the Gurobi software (applied to the QUBO formulation of these models). We found that the flexible PEPS exactly matches the energy for all the randomly generated problems and incurs only up to $1\%$ error for the QPLIB problems. Remarkably, our simple method was also able to find a \textit{better} solution for some of the QPLIB problems when the Gurobi optimization time was limited to 300 seconds, comparable to the time taken by the PEPS algorithm. These results demonstrate the potential of the flexible PEPS for solving hard combinatorial optimization problems.

\textit{(2) Simulated quantum annealing.} Next, we applied the flexible PEPS algorithm to simulate \textit{quantum annealing}, namely, to approximate the real-time dynamics of the time-dependent Hamiltonian
$
    H_{\mbox{\tiny SQA}}(\lambda(\tau)) = \lambda(\tau) H_{\mbox{\tiny Ising-SG}}-(1-\lambda(\tau))\sum_{i}\sigma^x_i
$, where $\sigma^x_i = \big(\begin{smallmatrix}
  0 & 1\\
  1 & 0
\end{smallmatrix}\big)$ is the Pauli $X$ matrix on site $i$, and $0 \leq \tau \leq 1$ is varied adiabatically between $\lambda(0) = 1$ and $\lambda(1) = 0$. We again chose the initial state $\ket{+}$, which here is the ground state of the initial Hamiltonian $H_{\mbox{\tiny SQA}}(0) = \sum_{i}\sigma^x_i$. When $\lambda(\tau)$ is varied adiabatically (see Appendix \ref{app:annealingschedule}), the time-evolved state $|\psi(\tau) \rangle \equiv e^{i\delta\tau H_{\mbox{\tiny SQA}}(\lambda(\tau))} |\psi\rangle$ remains close to the ground state of the instantaneous Hamiltonian, $H_{\mbox{\tiny SQA}}(\lambda(\tau))$, throughout the evolution, thus converging at $\lambda = 1$ to the ground state of the classical Ising model, $H_{\mbox{\tiny Ising-SG}}$ (equivalently, the solution of the corresponding QUBO). As opposed to the classical imaginary time-evolution discussed in the previous section, the evolution here is driven by a quantum Hamiltonian, which introduces a trotter error, though we mitigated this error by choosing a small trotter step. We used the annealing schedule plotted in Fig.~\ref{fig:sqa_schedule}, a re-scaled version of the schedule of the D-wave physical annealer. The results are shown in Fig.~\ref{fig:main_results}(top, left/right). We found that for randomly generated problems the algorithm performs poorly giving up to $75 \%$ error, but performed better on the structured QPLIB problems giving $\approx 8\%$ error. We believe that this poor performance is due to breaking of adiabaticity, which diverges the time-evolved state from the instantaneous ground state. As an attempt to course-correct, we also simulated a \textit{hybrid annealing}, which incorporates intermittent \textit{imaginary-time} evolution driven by the instantaneous Hamiltonian to steer the state back towards the instantaneous ground state. This approach gave much more accurate results, shown in Fig.~\ref{fig:main_results}(bottom), giving up to $4\%$ error for the randomly generated problems and up to $1\%$ error for QPLIB problems, comparable with classical imaginary time-evolution, but beating Gurobi on more problems.

\begin{figure}[t]
    \centering
    \includegraphics[width=\columnwidth]{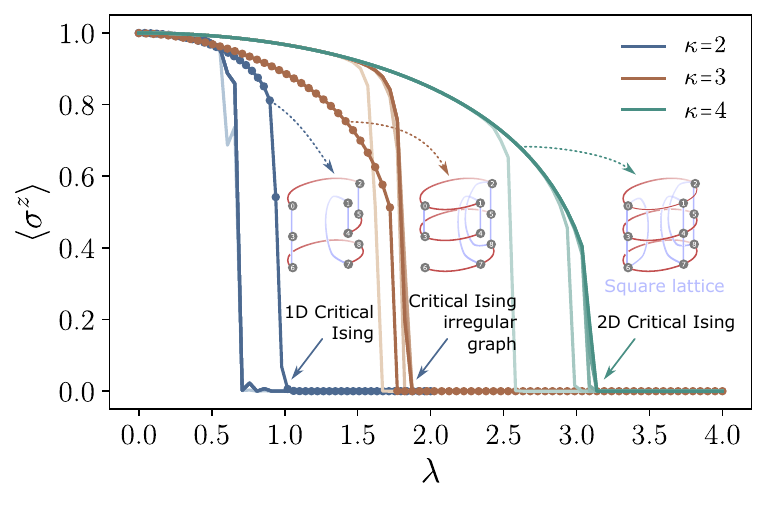}
    \caption{Average single site spin magnetization vs. the strength of the magnetic field $\lambda$ for the ground state of the uniform 2D quantum transverse field Ising model $H_{\mbox{\tiny TFIM}}$ on a square lattice. Curves without markers correspond to \textit{flexible} PEPS simulations with $\kappa = 2,3,4$. For each $\kappa$, the results were converged for increasing bond dimension, lighter curves correspond to smaller bond dimensions. The curves with markers correspond to \textit{fixed} PEPS simulations of Ising models defined on ``critical'' geometries (see also Appendix \ref{app:criticalPEPSgeometry}) that are recovered from the flexible-PEPS simulations.}
    \label{fig:TFIM_critical_comparison}
\end{figure}

(3) \textit{2D Uniform Quantum Transverse Field Ising Model.} We also applied the flexible PEPS to approximate the ground state of the \textit{uniform} quantum Ising model, $H_{\mbox{\tiny TFIM}} = \sum_i \sum_{ij} S_i^z S_j^z + \lambda \sum_i \sigma^x_i$, on an infinite square lattice. We approximated the ground states of this model using flexible PEPS with simple updates and a $3 \times 3$ unit cell, varying $\kappa = 2,3,4$. Fig.~\ref{fig:TFIM_critical_comparison} shows the single site magnetization averaged over the unit cell, computed using the ``simple measurements'' scheme, see Appendix \ref{app:measurements}. We found that PEPS with $\kappa = 4$, equal to the coordination number of the square lattice, accurately locates the 2D Ising critical point around $\approx 3.1$. But, as might perhaps be expected, flexible PEPS with $\kappa = 2,3$ performed poorly in locating the critical point. However, an interesting feature emerged -- the flexible PEPS curves for $\kappa = 2,3$ nonetheless displayed sharp transitions, at $\lambda \approx 1$ and $\lambda \approx 2$, respectively. (In fact, the average bond entanglement entropy also exhibited similar transitions, see Fig.~\ref{fig:TFIM_BEE_compare}.) In order to understand what these sharp transitions were, we tracked the geometry that the flexible PEPS acquires precisely at the transitions. These ``critical'' geometries for $\kappa = 2,3$ are depicted in Fig.~\ref{fig:TFIM_critical_comparison}. For $\kappa = 2$, we found that the critical geometry is in fact a chain with periodic boundaries, but with an optimization-determined site order. $\kappa = 3$ corresponded to a irregular graph with average vertex degree of $\approx 2.6$. (For $\kappa =4$, we recovered the 2D square lattice at the critical point.) We then simulated two different quantum Ising models using the usual \textit{fixed} PEPS with these respective geometries and found (solid markers in Fig.~\ref{fig:TFIM_critical_comparison}) that the locations of the sharp transitions, $\lambda \approx 1$ and $\lambda \approx 2$, coincide with the critical points of these two quantum Ising models. (In particular, $\lambda = 1 $ is the critical point of the quantum transverse field Ising model on a spin chain.) Thus, when the Hamiltonian is translation-invariant, there appears a correspondence between deleting PEPS bonds and truncating the corresponding Hamiltonian terms. This is compatible with the intuition that all PEPS bonds must be equally prominent when simulating translation-invariant models (i.e., $\mathcal{P} = \mathcal{G}$), which is realized here by our simple edge deletion rule (see also Fig.~\ref{fig:TFIM_BEE_compare}). Note that the geometry of the Flexible PEPS is changing along the solid curves, while the solid markers are for fixed PEPS based only on the critical geometries. Consequently, away from the critical points, for any fixed $\kappa$, the solid markers deviate from the solid curves. The question of whether the flexible PEPS simulation of a \textit{generic} translation-invariant model also traverses critical points of the same model in restricted geometries when tuning $\kappa$ will be tackled in future work.

\textit{Outlook.---} We have introduced a flexible PEPS algorithm that can adapt the underlying tensor network geometry to the correlation structure of the target state while maintaining a manageable number of bond indices, thus extending PEPS techniques to irregular and densely connected Hamiltonians (such as spin glasses and SYK models) made of hundreds of sites, even defined on fluctuating background geometries. We have demonstrated that a straightforward PEPS algorithm combining only simple updates with a bond entanglement entropy-based edge deletion rule already works remarkably well for a range of many-body problems, however, we expect to obtain a systematic improvement by replacing simple updates with cluster or full updates. For $\kappa = 2$, the PEPS reduces to a flexible MPS, namely, an MPS whose site order is adpatively fixed during the optimization. We found that the flexible MPS already worked well for some spin glass problems (see Fig.~\ref{fig:main_fig1}, Table.~\ref{tab:error_values}, and Appendix \ref{app:flexiblemps}) but its potential for simulating densely connected disordered models will be systematically assessed elsewhere.

\textit{Acknowledgments.--}
The simulations were partly run on the cluster at Donostia International Physics Center. We also acknowledge discussions with the technical and research teams of Multiverse Computing. 

\bibliography{references}

%apsrev4-2.bst 2019-01-14 (MD) hand-edited version of apsrev4-1.bst
%Control: key (0)
%Control: author (8) initials jnrlst
%Control: editor formatted (1) identically to author
%Control: production of article title (0) allowed
%Control: page (0) single
%Control: year (1) truncated
%Control: production of eprint (0) enabled
\begin{thebibliography}{28}%
\makeatletter
\providecommand \@ifxundefined [1]{%
 \@ifx{#1\undefined}
}%
\providecommand \@ifnum [1]{%
 \ifnum #1\expandafter \@firstoftwo
 \else \expandafter \@secondoftwo
 \fi
}%
\providecommand \@ifx [1]{%
 \ifx #1\expandafter \@firstoftwo
 \else \expandafter \@secondoftwo
 \fi
}%
\providecommand \natexlab [1]{#1}%
\providecommand \enquote  [1]{``#1''}%
\providecommand \bibnamefont  [1]{#1}%
\providecommand \bibfnamefont [1]{#1}%
\providecommand \citenamefont [1]{#1}%
\providecommand \href@noop [0]{\@secondoftwo}%
\providecommand \href [0]{\begingroup \@sanitize@url \@href}%
\providecommand \@href[1]{\@@startlink{#1}\@@href}%
\providecommand \@@href[1]{\endgroup#1\@@endlink}%
\providecommand \@sanitize@url [0]{\catcode `\\12\catcode `\$12\catcode `\&12\catcode `\#12\catcode `\^12\catcode `\_12\catcode `\%12\relax}%
\providecommand \@@startlink[1]{}%
\providecommand \@@endlink[0]{}%
\providecommand \url  [0]{\begingroup\@sanitize@url \@url }%
\providecommand \@url [1]{\endgroup\@href {#1}{\urlprefix }}%
\providecommand \urlprefix  [0]{URL }%
\providecommand \Eprint [0]{\href }%
\providecommand \doibase [0]{https://doi.org/}%
\providecommand \selectlanguage [0]{\@gobble}%
\providecommand \bibinfo  [0]{\@secondoftwo}%
\providecommand \bibfield  [0]{\@secondoftwo}%
\providecommand \translation [1]{[#1]}%
\providecommand \BibitemOpen [0]{}%
\providecommand \bibitemStop [0]{}%
\providecommand \bibitemNoStop [0]{.\EOS\space}%
\providecommand \EOS [0]{\spacefactor3000\relax}%
\providecommand \BibitemShut  [1]{\csname bibitem#1\endcsname}%
\let\auto@bib@innerbib\@empty
%</preamble>
\bibitem [{\citenamefont {Or\'us}(2014)}]{PEPSReviewOrus}%
  \BibitemOpen
  \bibfield  {author} {\bibinfo {author} {\bibfnamefont {R.}~\bibnamefont {Or\'us}},\ }\bibfield  {title} {\bibinfo {title} {A practical introduction to tensor networks: Matrix product states and projected entangled pair states},\ }\href {https://doi.org/doi.org/10.1016/j.aop.2014.06.013} {\bibfield  {journal} {\bibinfo  {journal} {Annals of Physics}\ }\textbf {\bibinfo {volume} {349}},\ \bibinfo {pages} {117} (\bibinfo {year} {2014})}\BibitemShut {NoStop}%
\bibitem [{\citenamefont {Cirac}\ \emph {et~al.}(2021)\citenamefont {Cirac}, \citenamefont {P\'erez-Garc\'{\i}a}, \citenamefont {Schuch},\ and\ \citenamefont {Verstraete}}]{PEPSReviewCirac}%
  \BibitemOpen
  \bibfield  {author} {\bibinfo {author} {\bibfnamefont {J.~I.}\ \bibnamefont {Cirac}}, \bibinfo {author} {\bibfnamefont {D.}~\bibnamefont {P\'erez-Garc\'{\i}a}}, \bibinfo {author} {\bibfnamefont {N.}~\bibnamefont {Schuch}},\ and\ \bibinfo {author} {\bibfnamefont {F.}~\bibnamefont {Verstraete}},\ }\bibfield  {title} {\bibinfo {title} {Matrix product states and projected entangled pair states: Concepts, symmetries, theorems},\ }\href {https://doi.org/10.1103/RevModPhys.93.045003} {\bibfield  {journal} {\bibinfo  {journal} {Rev. Mod. Phys.}\ }\textbf {\bibinfo {volume} {93}},\ \bibinfo {pages} {045003} (\bibinfo {year} {2021})}\BibitemShut {NoStop}%
\bibitem [{\citenamefont {Fannes}\ \emph {et~al.}(1992)\citenamefont {Fannes}, \citenamefont {Nachtergaele},\ and\ \citenamefont {Werner}}]{Fannes}%
  \BibitemOpen
  \bibfield  {author} {\bibinfo {author} {\bibfnamefont {M.}~\bibnamefont {Fannes}}, \bibinfo {author} {\bibfnamefont {B.}~\bibnamefont {Nachtergaele}},\ and\ \bibinfo {author} {\bibfnamefont {R.~F.}\ \bibnamefont {Werner}},\ }\bibfield  {title} {\bibinfo {title} {Finitely correlated states on quantum spin chains},\ }\href@noop {} {\bibfield  {journal} {\bibinfo  {journal} {Communications in Mathematical Physics}\ }\textbf {\bibinfo {volume} {144}},\ \bibinfo {pages} {443 } (\bibinfo {year} {1992})}\BibitemShut {NoStop}%
\bibitem [{\citenamefont {Verstraete}\ and\ \citenamefont {Cirac}(2004)}]{PEPS}%
  \BibitemOpen
  \bibfield  {author} {\bibinfo {author} {\bibfnamefont {F.}~\bibnamefont {Verstraete}}\ and\ \bibinfo {author} {\bibfnamefont {J.~I.}\ \bibnamefont {Cirac}},\ }\href {https://arxiv.org/abs/cond-mat/0407066} {\bibinfo {title} {Renormalization algorithms for quantum-many body systems in two and higher dimensions}} (\bibinfo {year} {2004}),\ \Eprint {https://arxiv.org/abs/cond-mat/0407066} {arXiv:cond-mat/0407066 [cond-mat.str-el]} \BibitemShut {NoStop}%
\bibitem [{\citenamefont {Jahromi}\ and\ \citenamefont {Or\'us}(2019)}]{gPEPS}%
  \BibitemOpen
  \bibfield  {author} {\bibinfo {author} {\bibfnamefont {S.~S.}\ \bibnamefont {Jahromi}}\ and\ \bibinfo {author} {\bibfnamefont {R.}~\bibnamefont {Or\'us}},\ }\bibfield  {title} {\bibinfo {title} {Universal tensor-network algorithm for any infinite lattice},\ }\href {https://doi.org/10.1103/PhysRevB.99.195105} {\bibfield  {journal} {\bibinfo  {journal} {Phys. Rev. B}\ }\textbf {\bibinfo {volume} {99}},\ \bibinfo {pages} {195105} (\bibinfo {year} {2019})}\BibitemShut {NoStop}%
\bibitem [{\citenamefont {Schuch}\ \emph {et~al.}(2011)\citenamefont {Schuch}, \citenamefont {P\'erez-Garc\'{\i}a},\ and\ \citenamefont {Cirac}}]{PEPSClassifyingPhases}%
  \BibitemOpen
  \bibfield  {author} {\bibinfo {author} {\bibfnamefont {N.}~\bibnamefont {Schuch}}, \bibinfo {author} {\bibfnamefont {D.}~\bibnamefont {P\'erez-Garc\'{\i}a}},\ and\ \bibinfo {author} {\bibfnamefont {I.}~\bibnamefont {Cirac}},\ }\bibfield  {title} {\bibinfo {title} {Classifying quantum phases using matrix product states and projected entangled pair states},\ }\href {https://doi.org/10.1103/PhysRevB.84.165139} {\bibfield  {journal} {\bibinfo  {journal} {Phys. Rev. B}\ }\textbf {\bibinfo {volume} {84}},\ \bibinfo {pages} {165139} (\bibinfo {year} {2011})}\BibitemShut {NoStop}%
\bibitem [{\citenamefont {Schuch}\ \emph {et~al.}(2013)\citenamefont {Schuch}, \citenamefont {Poilblanc}, \citenamefont {Cirac},\ and\ \citenamefont {P\'erez-Garc\'{\i}a}}]{PEPSTopologicalOrder}%
  \BibitemOpen
  \bibfield  {author} {\bibinfo {author} {\bibfnamefont {N.}~\bibnamefont {Schuch}}, \bibinfo {author} {\bibfnamefont {D.}~\bibnamefont {Poilblanc}}, \bibinfo {author} {\bibfnamefont {J.~I.}\ \bibnamefont {Cirac}},\ and\ \bibinfo {author} {\bibfnamefont {D.}~\bibnamefont {P\'erez-Garc\'{\i}a}},\ }\bibfield  {title} {\bibinfo {title} {Topological order in the projected entangled-pair states formalism: Transfer operator and boundary hamiltonians},\ }\href {https://doi.org/10.1103/PhysRevLett.111.090501} {\bibfield  {journal} {\bibinfo  {journal} {Phys. Rev. Lett.}\ }\textbf {\bibinfo {volume} {111}},\ \bibinfo {pages} {090501} (\bibinfo {year} {2013})}\BibitemShut {NoStop}%
\bibitem [{\citenamefont {Poilblanc}\ \emph {et~al.}(2012)\citenamefont {Poilblanc}, \citenamefont {Schuch}, \citenamefont {P\'erez-Garc\'{\i}a},\ and\ \citenamefont {Cirac}}]{PEPSSpinLiquid}%
  \BibitemOpen
  \bibfield  {author} {\bibinfo {author} {\bibfnamefont {D.}~\bibnamefont {Poilblanc}}, \bibinfo {author} {\bibfnamefont {N.}~\bibnamefont {Schuch}}, \bibinfo {author} {\bibfnamefont {D.}~\bibnamefont {P\'erez-Garc\'{\i}a}},\ and\ \bibinfo {author} {\bibfnamefont {J.~I.}\ \bibnamefont {Cirac}},\ }\bibfield  {title} {\bibinfo {title} {Topological and entanglement properties of resonating valence bond wave functions},\ }\href {https://doi.org/10.1103/PhysRevB.86.014404} {\bibfield  {journal} {\bibinfo  {journal} {Phys. Rev. B}\ }\textbf {\bibinfo {volume} {86}},\ \bibinfo {pages} {014404} (\bibinfo {year} {2012})}\BibitemShut {NoStop}%
\bibitem [{\citenamefont {van Hemmen}(1982)}]{SpinGlass1}%
  \BibitemOpen
  \bibfield  {author} {\bibinfo {author} {\bibfnamefont {J.~L.}\ \bibnamefont {van Hemmen}},\ }\bibfield  {title} {\bibinfo {title} {Classical spin-glass model},\ }\href {https://doi.org/10.1103/PhysRevLett.49.409} {\bibfield  {journal} {\bibinfo  {journal} {Phys. Rev. Lett.}\ }\textbf {\bibinfo {volume} {49}},\ \bibinfo {pages} {409} (\bibinfo {year} {1982})}\BibitemShut {NoStop}%
\bibitem [{\citenamefont {Binder}\ and\ \citenamefont {Young}(1986)}]{SpinGlass2}%
  \BibitemOpen
  \bibfield  {author} {\bibinfo {author} {\bibfnamefont {K.}~\bibnamefont {Binder}}\ and\ \bibinfo {author} {\bibfnamefont {A.~P.}\ \bibnamefont {Young}},\ }\bibfield  {title} {\bibinfo {title} {Spin glasses: Experimental facts, theoretical concepts, and open questions},\ }\href {https://doi.org/10.1103/RevModPhys.58.801} {\bibfield  {journal} {\bibinfo  {journal} {Rev. Mod. Phys.}\ }\textbf {\bibinfo {volume} {58}},\ \bibinfo {pages} {801} (\bibinfo {year} {1986})}\BibitemShut {NoStop}%
\bibitem [{\citenamefont {Mezard}\ \emph {et~al.}(1987)\citenamefont {Mezard}, \citenamefont {Parisi},\ and\ \citenamefont {Virasoro}}]{SpinGlassBook}%
  \BibitemOpen
  \bibfield  {author} {\bibinfo {author} {\bibfnamefont {M.}~\bibnamefont {Mezard}}, \bibinfo {author} {\bibfnamefont {G.}~\bibnamefont {Parisi}},\ and\ \bibinfo {author} {\bibfnamefont {M.}~\bibnamefont {Virasoro}},\ }\href {https://books.google.ca/books?id=DwY8DQAAQBAJ} {\emph {\bibinfo {title} {Spin Glass Theory And Beyond: An Introduction To The Replica Method And Its Applications}}},\ World Scientific Lecture Notes In Physics\ (\bibinfo  {publisher} {World Scientific Publishing Company},\ \bibinfo {year} {1987})\BibitemShut {NoStop}%
\bibitem [{\citenamefont {Lucas}(2014)}]{SpinGlassCombinatorialOptimization}%
  \BibitemOpen
  \bibfield  {author} {\bibinfo {author} {\bibfnamefont {A.}~\bibnamefont {Lucas}},\ }\bibfield  {title} {\bibinfo {title} {Ising formulations of many np problems},\ }\bibfield  {journal} {\bibinfo  {journal} {Frontiers in Physics}\ }\textbf {\bibinfo {volume} {2}},\ \href {https://doi.org/10.3389/fphy.2014.00005} {10.3389/fphy.2014.00005} (\bibinfo {year} {2014})\BibitemShut {NoStop}%
\bibitem [{\citenamefont {Sachdev}\ and\ \citenamefont {Ye}(1993)}]{SYK1}%
  \BibitemOpen
  \bibfield  {author} {\bibinfo {author} {\bibfnamefont {S.}~\bibnamefont {Sachdev}}\ and\ \bibinfo {author} {\bibfnamefont {J.}~\bibnamefont {Ye}},\ }\bibfield  {title} {\bibinfo {title} {Gapless spin-fluid ground state in a random quantum heisenberg magnet},\ }\href {https://doi.org/10.1103/PhysRevLett.70.3339} {\bibfield  {journal} {\bibinfo  {journal} {Phys. Rev. Lett.}\ }\textbf {\bibinfo {volume} {70}},\ \bibinfo {pages} {3339} (\bibinfo {year} {1993})}\BibitemShut {NoStop}%
\bibitem [{\citenamefont {Kitaev}()}]{SYK2}%
  \BibitemOpen
  \bibfield  {author} {\bibinfo {author} {\bibfnamefont {A.}~\bibnamefont {Kitaev}},\ }\href@noop {} {\bibinfo {title} {A simple model of quantum holography, talks at kitp, april 7, 2015 and may 27, 2015.}}\BibitemShut {Stop}%
\bibitem [{\citenamefont {Chan}(2024)}]{QuantumChemistryReview}%
  \BibitemOpen
  \bibfield  {author} {\bibinfo {author} {\bibfnamefont {G.~K.-L.}\ \bibnamefont {Chan}},\ }\bibfield  {title} {\bibinfo {title} {Quantum chemistry{,} classical heuristics{,} and quantum advantage},\ }\href {https://doi.org/10.1039/D4FD00141A} {\bibfield  {journal} {\bibinfo  {journal} {Faraday Discuss.}\ ,\ } (\bibinfo {year} {2024})}\BibitemShut {NoStop}%
\bibitem [{\citenamefont {Szalay}\ \emph {et~al.}(2015)\citenamefont {Szalay}, \citenamefont {Pfeffer}, \citenamefont {Murg}, \citenamefont {Barcza}, \citenamefont {Verstraete}, \citenamefont {Schneider},\ and\ \citenamefont {Legeza}}]{QuantumChemistry}%
  \BibitemOpen
  \bibfield  {author} {\bibinfo {author} {\bibfnamefont {S.}~\bibnamefont {Szalay}}, \bibinfo {author} {\bibfnamefont {M.}~\bibnamefont {Pfeffer}}, \bibinfo {author} {\bibfnamefont {V.}~\bibnamefont {Murg}}, \bibinfo {author} {\bibfnamefont {G.}~\bibnamefont {Barcza}}, \bibinfo {author} {\bibfnamefont {F.}~\bibnamefont {Verstraete}}, \bibinfo {author} {\bibfnamefont {R.}~\bibnamefont {Schneider}},\ and\ \bibinfo {author} {\bibfnamefont {O.}~\bibnamefont {Legeza}},\ }\bibfield  {title} {\bibinfo {title} {Tensor product methods and entanglement optimization for ab initio quantum chemistry},\ }\href {https://doi.org/https://doi.org/10.1002/qua.24898} {\bibfield  {journal} {\bibinfo  {journal} {International Journal of Quantum Chemistry}\ }\textbf {\bibinfo {volume} {115}},\ \bibinfo {pages} {1342} (\bibinfo {year} {2015})},\ \Eprint {https://arxiv.org/abs/https://onlinelibrary.wiley.com/doi/pdf/10.1002/qua.24898} {https://onlinelibrary.wiley.com/doi/pdf/10.1002/qua.24898} \BibitemShut {NoStop}%
\bibitem [{\citenamefont {Patra}\ \emph {et~al.}(2024)\citenamefont {Patra}, \citenamefont {Jahromi}, \citenamefont {Singh},\ and\ \citenamefont {Or\'us}}]{ibm_simulator}%
  \BibitemOpen
  \bibfield  {author} {\bibinfo {author} {\bibfnamefont {S.}~\bibnamefont {Patra}}, \bibinfo {author} {\bibfnamefont {S.~S.}\ \bibnamefont {Jahromi}}, \bibinfo {author} {\bibfnamefont {S.}~\bibnamefont {Singh}},\ and\ \bibinfo {author} {\bibfnamefont {R.}~\bibnamefont {Or\'us}},\ }\bibfield  {title} {\bibinfo {title} {Efficient tensor network simulation of ibm's largest quantum processors},\ }\href {https://doi.org/10.1103/PhysRevResearch.6.013326} {\bibfield  {journal} {\bibinfo  {journal} {Phys. Rev. Res.}\ }\textbf {\bibinfo {volume} {6}},\ \bibinfo {pages} {013326} (\bibinfo {year} {2024})}\BibitemShut {NoStop}%
\bibitem [{\citenamefont {Jiang}\ \emph {et~al.}(2008)\citenamefont {Jiang}, \citenamefont {Weng},\ and\ \citenamefont {Xiang}}]{SimpleUpdate}%
  \BibitemOpen
  \bibfield  {author} {\bibinfo {author} {\bibfnamefont {H.~C.}\ \bibnamefont {Jiang}}, \bibinfo {author} {\bibfnamefont {Z.~Y.}\ \bibnamefont {Weng}},\ and\ \bibinfo {author} {\bibfnamefont {T.}~\bibnamefont {Xiang}},\ }\bibfield  {title} {\bibinfo {title} {Accurate determination of tensor network state of quantum lattice models in two dimensions},\ }\href {https://doi.org/10.1103/PhysRevLett.101.090603} {\bibfield  {journal} {\bibinfo  {journal} {Phys. Rev. Lett.}\ }\textbf {\bibinfo {volume} {101}},\ \bibinfo {pages} {090603} (\bibinfo {year} {2008})}\BibitemShut {NoStop}%
\bibitem [{\citenamefont {Lubasch}\ \emph {et~al.}(2014{\natexlab{a}})\citenamefont {Lubasch}, \citenamefont {Cirac},\ and\ \citenamefont {Ba\~nuls}}]{PEPSUpdates1}%
  \BibitemOpen
  \bibfield  {author} {\bibinfo {author} {\bibfnamefont {M.}~\bibnamefont {Lubasch}}, \bibinfo {author} {\bibfnamefont {J.~I.}\ \bibnamefont {Cirac}},\ and\ \bibinfo {author} {\bibfnamefont {M.-C.}\ \bibnamefont {Ba\~nuls}},\ }\bibfield  {title} {\bibinfo {title} {Algorithms for finite projected entangled pair states},\ }\href {https://doi.org/10.1103/PhysRevB.90.064425} {\bibfield  {journal} {\bibinfo  {journal} {Phys. Rev. B}\ }\textbf {\bibinfo {volume} {90}},\ \bibinfo {pages} {064425} (\bibinfo {year} {2014}{\natexlab{a}})}\BibitemShut {NoStop}%
\bibitem [{\citenamefont {Lubasch}\ \emph {et~al.}(2014{\natexlab{b}})\citenamefont {Lubasch}, \citenamefont {Cirac},\ and\ \citenamefont {Bañuls}}]{PEPSUpdates2}%
  \BibitemOpen
  \bibfield  {author} {\bibinfo {author} {\bibfnamefont {M.}~\bibnamefont {Lubasch}}, \bibinfo {author} {\bibfnamefont {J.~I.}\ \bibnamefont {Cirac}},\ and\ \bibinfo {author} {\bibfnamefont {M.-C.}\ \bibnamefont {Bañuls}},\ }\bibfield  {title} {\bibinfo {title} {Unifying projected entangled pair state contractions},\ }\href {https://doi.org/10.1088/1367-2630/16/3/033014} {\bibfield  {journal} {\bibinfo  {journal} {New Journal of Physics}\ }\textbf {\bibinfo {volume} {16}},\ \bibinfo {pages} {033014} (\bibinfo {year} {2014}{\natexlab{b}})}\BibitemShut {NoStop}%
\bibitem [{Note1()}]{Note1}%
  \BibitemOpen
  \bibinfo {note} {For concreteness, we focus on PEPS time-evolution algorithms to approximate the low-energy states and dynamics of the system, but our flexible PEPS updates can also be incorporated in PEPS-based variational algorithms}\BibitemShut {NoStop}%
\bibitem [{Note2()}]{Note2}%
  \BibitemOpen
  \bibinfo {note} {Consider a graph with 6 nodes numbered 0,1,...,5, and edges: [(0,1), (1,2), (2,3), (3,4), (4,5), (0,3), (2,4), (3,5)]. We can group these edges into four layers of non-intersecting edges: Layer1=[(0,1), (2,3), (4,5)] Layer2=[(1,2), (3,4)] Layer3=[(0,3), (2,4)] Layer4=[(3,5)]. Note that for several problems studied in this paper we also tested the straightforward gate sequence $(0,1),(0,2)...(1,2),(1,3)...$ with with a small trotter step $= 0.01$. Surprisingly, we obtained comparable results despite incurring an uncontrolled trotter error.}\BibitemShut {Stop}%
\bibitem [{Note3()}]{Note3}%
  \BibitemOpen
  \bibinfo {note} {We also tried a slightly different edge-deletion strategy. If the new bond resulting from the current gate update has bond entanglement in the lowest $30\%$ then we remove that bond. Otherwise, we remove the bond with the least bond entanglement entropy. This strategy worked better for some models.}\BibitemShut {Stop}%
\bibitem [{\citenamefont {Barahona}(1982)}]{SpinGlassNPHard}%
  \BibitemOpen
  \bibfield  {author} {\bibinfo {author} {\bibfnamefont {F.}~\bibnamefont {Barahona}},\ }\bibfield  {title} {\bibinfo {title} {On the computational complexity of ising spin glass models},\ }\href {https://doi.org/10.1088/0305-4470/15/10/028} {\bibfield  {journal} {\bibinfo  {journal} {Journal of Physics A: Mathematical and General}\ }\textbf {\bibinfo {volume} {15}},\ \bibinfo {pages} {3241} (\bibinfo {year} {1982})}\BibitemShut {NoStop}%
\bibitem [{\citenamefont {Stoudenmire}\ and\ \citenamefont {White}(2010)}]{MPSSampling1}%
  \BibitemOpen
  \bibfield  {author} {\bibinfo {author} {\bibfnamefont {E.~M.}\ \bibnamefont {Stoudenmire}}\ and\ \bibinfo {author} {\bibfnamefont {S.~R.}\ \bibnamefont {White}},\ }\bibfield  {title} {\bibinfo {title} {Minimally entangled typical thermal state algorithms},\ }\href {https://doi.org/10.1088/1367-2630/12/5/055026} {\bibfield  {journal} {\bibinfo  {journal} {New Journal of Physics}\ }\textbf {\bibinfo {volume} {12}},\ \bibinfo {pages} {055026} (\bibinfo {year} {2010})}\BibitemShut {NoStop}%
\bibitem [{\citenamefont {Ferris}\ and\ \citenamefont {Vidal}(2012)}]{MPSSampling2}%
  \BibitemOpen
  \bibfield  {author} {\bibinfo {author} {\bibfnamefont {A.~J.}\ \bibnamefont {Ferris}}\ and\ \bibinfo {author} {\bibfnamefont {G.}~\bibnamefont {Vidal}},\ }\bibfield  {title} {\bibinfo {title} {Perfect sampling with unitary tensor networks},\ }\href {https://doi.org/10.1103/PhysRevB.85.165146} {\bibfield  {journal} {\bibinfo  {journal} {Phys. Rev. B}\ }\textbf {\bibinfo {volume} {85}},\ \bibinfo {pages} {165146} (\bibinfo {year} {2012})}\BibitemShut {NoStop}%
\bibitem [{\citenamefont {Furini}\ \emph {et~al.}(2019)\citenamefont {Furini}, \citenamefont {Traversi}, \citenamefont {Belotti}, \citenamefont {Frangioni}, \citenamefont {Gleixner}, \citenamefont {Gould}, \citenamefont {Liberti}, \citenamefont {Lodi}, \citenamefont {Misener}, \citenamefont {Mittelmann}, \citenamefont {Sahinidis}, \citenamefont {Vigerske},\ and\ \citenamefont {Wiegele}}]{QPLIB}%
  \BibitemOpen
  \bibfield  {author} {\bibinfo {author} {\bibfnamefont {F.}~\bibnamefont {Furini}}, \bibinfo {author} {\bibfnamefont {E.}~\bibnamefont {Traversi}}, \bibinfo {author} {\bibfnamefont {P.}~\bibnamefont {Belotti}}, \bibinfo {author} {\bibfnamefont {A.}~\bibnamefont {Frangioni}}, \bibinfo {author} {\bibfnamefont {A.~M.}\ \bibnamefont {Gleixner}}, \bibinfo {author} {\bibfnamefont {N.}~\bibnamefont {Gould}}, \bibinfo {author} {\bibfnamefont {L.}~\bibnamefont {Liberti}}, \bibinfo {author} {\bibfnamefont {A.}~\bibnamefont {Lodi}}, \bibinfo {author} {\bibfnamefont {R.}~\bibnamefont {Misener}}, \bibinfo {author} {\bibfnamefont {H.~D.}\ \bibnamefont {Mittelmann}}, \bibinfo {author} {\bibfnamefont {N.~V.}\ \bibnamefont {Sahinidis}}, \bibinfo {author} {\bibfnamefont {S.}~\bibnamefont {Vigerske}},\ and\ \bibinfo {author} {\bibfnamefont {A.}~\bibnamefont {Wiegele}},\ }\bibfield  {title} {\bibinfo {title} {{QPLIB:} a library of quadratic programming instances},\ }\href {https://doi.org/10.1007/S12532-018-0147-4} {\bibfield
  {journal} {\bibinfo  {journal} {Math. Program. Comput.}\ }\textbf {\bibinfo {volume} {11}},\ \bibinfo {pages} {237} (\bibinfo {year} {2019})}\BibitemShut {NoStop}%
\bibitem [{Note4()}]{Note4}%
  \BibitemOpen
  \bibinfo {note} {The MIPGap returned by Gurobi estimates an upper bound on the error in the solution.}\BibitemShut {Stop}%
\end{thebibliography}%

\appendix
%\tableofcontents

\section{PEPS update algorithms}\label{app:pepsalgorithms}
In this appendix, we briefly review how the PEPS is updated after applying a two-site time-evolution gate, $O_{ij}$, on sites $i$ and $j$, resulting in new values of the vertex tensors $\Gamma^{[i]}$ and $\Gamma^{[j]}$ and the bond tensor $\lambda^{[i,j]}$. The simplest scheme is to update these tensors locally and ignoring the influence of the remaining PEPS tensors. This \textit{simple updates} scheme often works surprisingly well. However, it can be systematically improved by applying \textit{cluster updates}, where a small cluster of neighbouring tensors around the sites $i$ and $j$ are included when updating the PEPS, or by \textit{full updates}, where all the PEPS tensors are included when updating the PEPS after absorbing any two-site gate.  Cluster and full-updates are generally more accurate than simple updates, but require a more sophisticated code implementation and incur a much higher computational cost. 

The same consideration also applies when computing the expectation value of local observables from a PEPS: one can either ignore the PEPS environment tensors around the local observable, or include a small cluster of them, or include all the PEPS tensors, which we refer to as the \textit{simple measurement}, \textit{cluster measurement}, and \textit{full measurement} scheme, respectively.

\subsection{Simple update}\label{app:simpleupdate}
\begin{figure}[t]
    \centering
    \includegraphics[width=0.48\textwidth]{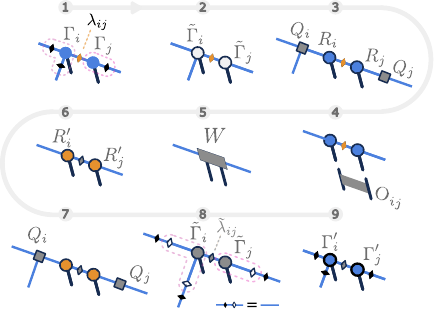}
    \caption{\textit{Step $1 \rightarrow 2$}. Multiply all the bond tensors into $\Gamma_i$ and $\Gamma_j$ as grouped by the dashed contours to obtain $\tilde{\Gamma}_i, \tilde{\Gamma}_j$ respectively. \textit{Step $2 \rightarrow 3$}. Perform QR decomposition of the tensors $\tilde{\Gamma}_i = Q_i R_i, \tilde{\Gamma}_j = R_j Q_j$ where the $Q$'s collects out all the virtual indices in the diagram, except $(i,j)$, and $R$'s retains the two physical indices. \textit{Steps $3 \rightarrow 5$}. Multiply the two-site gate $O_{ij}$ with tensors $R_i$ and $R_j$ to obtain a 4-index tensor $W$. \textit{Step $5 \rightarrow 6$}. Singular value decompose $W$ to obtain tensors $R'_i,R'_j$ and the updated bond tensor $\lambda'_{i,j}$. \textit{Step $6 \rightarrow 7$}. Multiple $R'_i$ with $Q_i$ and $R'_j$ with $Q_j$ to obtain tensors $\tilde{\Gamma}'_i$, and  $\tilde{\Gamma}'_j$ respectively. \textit{Step $7 \rightarrow 8$}. Multiple the inverse of the bond tensors with  $\tilde{\Gamma}'_i$ and $\tilde{\Gamma}'_j$, according to the dashed contour grouping, to obtain the updated PEPS tensors $\Gamma'_i$, and $\Gamma'_j$.}
    \label{fig:simpleupdate}
\end{figure}

Fig.\ref{fig:simpleupdate} depicts the steps of the simple update, which basically proceeds as follows: (i) The vertex tensors $\Gamma^{[i]}$,  $\Gamma^{[j]}$ are compressed into $\tilde{\Gamma}^{[i]}$,  $\tilde{\Gamma}^{[j]}$ by means of QR decompositions, (ii) the gate is multiplied with the compressed vertex tensors and the bond tensor $\lambda^{[i,j]}$, (iii) the resulting tensor is suitably reshaped into a matrix and factorized by singular value decomposition, (iv) only the largest $\chi_{i,j}$ number of singular values (and the corresponding left and right singular vectors) are preserved while the remaining are discarded, and (v) the truncated factors are suitably reshaped to obtained the updated vertex tensors $\tilde{\Gamma}^{[i]}$,  $\tilde{\Gamma}^{[j]}$.

\begin{figure}[t]
    \centering
    \includegraphics[width=0.37\textwidth]{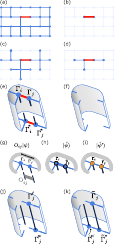}
    \caption{(a) An PEPS update is indicated for a gate applied between sites $i$ and $j$. (Physical indices are suppressed in panels (a)-(d)). (b) Simple update corresponds to ignoring all remaining tensors in the update. (c) A cluster update scheme that includes neighbouring tensors, here shown, neighbours located at most two edges away from the sites $i,j$. (d) A cluster update scheme that includes only first neighbours. (e) The tensor network expression for the reduced density matrix $\rho$ of sites $i,j$. (e) The environment tensor for the reduced density matrix $\rho$. (g)-(i) Absorbing a two-site gate $O_{ij}$ into the PEPS. (j)-(k) Partial environment tensors involved in the update, see text.}
    \label{fig:clusterupdate}
\end{figure}

\begin{figure}[t]
    \centering
    \includegraphics[width=\columnwidth]{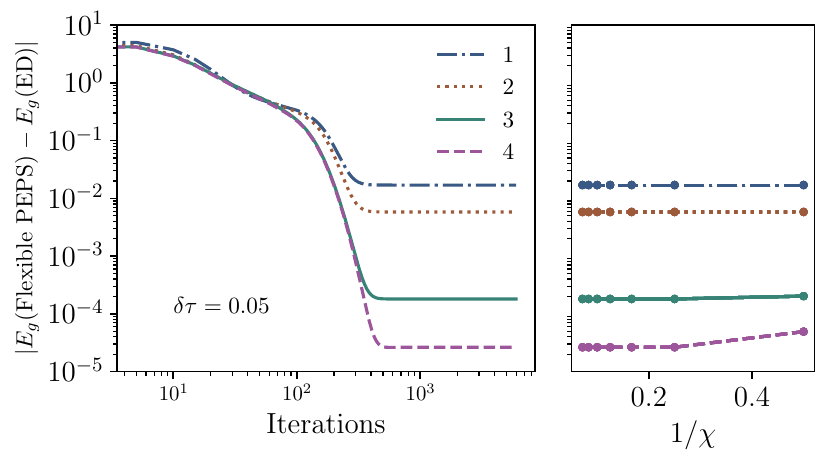}
    \caption{Comparison with exact diagonalization result: random TFIM on 2D square lattice of dimension $4\times 3$. The ground state is obtained by doing TEBD with a fixed imaginary time step $\beta=0.05$. The left panel shows the absolute difference between energy obtained from local TN contraction and exact diagonalization, where we present results obtained with $NVB=1,2,3,4$ with bond dimension $\chi=14$. The right panel shows the bond dimension scaling of the same absolute difference.}
    \label{fig:rTFIM_12}
\end{figure}

\textit{Accuracy of simple updates.} In order to assess the accuracy of simple updates for our purposes, we used the flexible PEPS based on simple updates to approximate the ground state of a small random 2D quantum Ising model,
\begin{eqnarray}\label{eq:disorder_ising}
    H_{\mbox{\tiny rand-ising}} = \sum_i h_i \sigma^z_i + \sum_{ij} J_{ij} S_i^z S_j^z + \sum_i g_i \sigma^x_i,
\end{eqnarray}
defined on a $4 \times 3$ lattice. Couplings $h_i, J_{ij}, g_i$ were chosen randomly from an uniform distribution. Fig. \ref{fig:rTFIM_12}  shows the error in ground state energy. (We studied several disorder instances of this model. The plot shows the result for a particular instance but we found a similar result for all the instances that we considered.) We approximated the ground state using imaginary-time evolution employing the flexible PEPS algorithm varying $\kappa = 2,3,4$, while the exact energy was obtained using diagonalization. We found that the accuracy of the simulation increased with $\kappa$. However, even for $\kappa = 4$ the error was significant, $\approx 10^{-5}$. At $\kappa = 4$, which equals the coordination number of the lattice, there are no edge deletions, suggesting that the main source of error are the \textit{simple} updates and measurements. (We used a small trotter step to mitigate the trotter error.) These results highlight the limitation of the simple updates + simple measurement scheme, which we have used in all the numerical experiments reported in this paper. Therefore, we expect that the results presented in this paper using simple updates -- evidently very accurate for some models -- can be systematically improved by using cluster and full updates.

\subsection{Cluster and Full updates}\label{app:clusterfull}
    
In contrast to simple updates, cluster and full update schemes include environment tensors when absorbing for a two-site gate, see Fig.~\ref{fig:clusterupdate}(a)-(d). A full update includes all the PEPS tensors as the environment when absorbing a single gate, panel (a). In extreme contrast, the simple update ignores all the remaining PEPS tensors, panel (b). For a cluster update, we include a small set of neighbouring tensors as the environment, for instance, tensors located at most two, panel (c), or at most one edges, panel (d), away from sites $i$ and $j$. Thus, the cluster and full updates differ from the simple update in the presence of the environment tensor,  see Fig.~\ref{fig:clusterupdate} (e)-(f). 

The cluster and full updates proceed as follows. First, keeping aside the environment tensor, we follow all the steps of the simple update procedure prior to the truncation of singular values to obtain the modified vertex tensors $\tilde{\Gamma}_i$ and $\tilde{\Gamma}_j$. The straightforward truncation step of the simple update is now replaced with a variational optimization involving the environment tensor. Let $E$ denote the \textit{ket} environment tensor, obtained by contracting together either a cluster of neighbouring PEPS tensors or all remaining PEPS tensors. $E^\dagger$ is the corresponding the \textit{bra} environment tensor, and $E E^\dagger$ is the total environment tensor for the two-site reduced density matrix. The product of $\tilde{\Gamma}_i$, $\tilde{\Gamma}_j$, and the total ket environment $E$ gives a tensor that can be understood as a quantum many-body state $\ket{\tilde{\psi}}$. If the new bond requires truncation i.e., if it's dimension is larger than the set maximum bond dimension, truncate the singular values and the $\Gamma$ tensors -- as is done in the simple update -- to obtain the truncated tensors ${\Gamma}'_i$ and ${\Gamma}'_j$. (The singular values are absorbed into the $\Gamma'$ tensors.) The product of the truncated tensors ${\Gamma}'_i$, ${\Gamma}'_j$, and $E$ then defines another state $\ket{\psi'}$. Next, variationally optimize the tensors ${\Gamma}'_i, {\Gamma}'_j$ to minimize the distance
    \begin{eqnarray}
         d(\tilde{\psi},{\psi}') &=& ||~ |\tilde{\psi} \rangle- |{\psi}' \rangle  ~||^2 \\ \nonumber
         &=& \langle \tilde{\psi} | \tilde{\psi} \rangle + \langle \psi | \psi \rangle- \langle \tilde{\psi} | \psi \rangle - \langle \psi | \tilde{\psi} \rangle
         \label{eq:variational_cost}
    \end{eqnarray}
by optimizing one vertex tensor at a time. First, we fix $\Gamma'_j$ and try to optimize $\Gamma'_i$, which translates the cost function to  
        \begin{eqnarray}
        d(\tilde{\psi},{\psi}') = \langle \tilde{\psi} | \tilde{\psi} \rangle+ {\Gamma'}_i^{\dagger} A \Gamma'_i - {\Gamma'}_i^{\dagger} B - B^{\dagger} {\Gamma'}_i,
        \label{eq:variational_cost_individual}
        \end{eqnarray}
where tensors $A, B$ are shown in Fig.\ref{fig:clusterupdate}(j),(k), respectively
The optimal solution is $\Gamma'_i=A^{-1} B$. After updating $\Gamma'$ as such, we proceed to optimize $\Gamma'_j$ similarly. These steps are iterated until tensors $\Gamma'_i$ and $\Gamma'_j$ converge.

\subsection{Computing PEPS expectation values} \label{app:measurements}
We briefly review how expectation values are obtained from the PEPS, which may have been obtained from an optimization based on simple, cluster, or full updates. Fig.~\ref{fig:clusterupdate}(e) depicts the most general tensor network expression for the local density matrix that inputs into the expectation value calculation. If the environment tensor is obtained by contracting all the remaining PEPS tensors, we'll call the measurement 
a \textit{full} measurement. Similarly, if the environment tensor is obtained by contracting only a cluster of neighbouring tensors, we'll refer to the the measurement as a \textit{cluster} measurement. A \textit{simple} measurement corresponds to replacing the environment tensor with the Identity.

All the PEPS results presented in this paper use simple updates + simple measurements. Combining simple updates with cluster measurements did not seem to improve the accuracy. However, we expect the accuracy to generally improve when using cluster and full updates, especially when implemented together with cluster or full measurements.

\section{Sampling from the flexible PEPS}\label{app:sample}
When applied to approximate the ground state of a classical Hamiltonian, such as \eqref{eq:spinglass}, our flexible PEPS algorithm often converged to an entangled PEPS with a small bond dimension (= 2 or 3) instead of a product state. Nonetheless, if the simulation is carried out reliably, the converged entangled state is expected to have a large overlap with the ground product state. In this case, we can hope to recover the latter by \textit{sampling} from the converged PEPS. 

The brute force sampling approach would be to contract the PEPS together and recover the probabilities of all the basis (product) states and then sample a product state according to that probability distribution, incurring a computational cost that scales exponentially with the size of the PEPS. Below, we describe an efficient PEPS sampling method that does not require contracting the whole tensor network. 
To this end, we adapted the sampling algorithm for the MPS \cite{MPSSampling1, MPSSampling2} to the PEPS tensor network based on an arbitrary geometry. 

\begin{figure*}[t]
    \centering
    \includegraphics[width=1\textwidth]{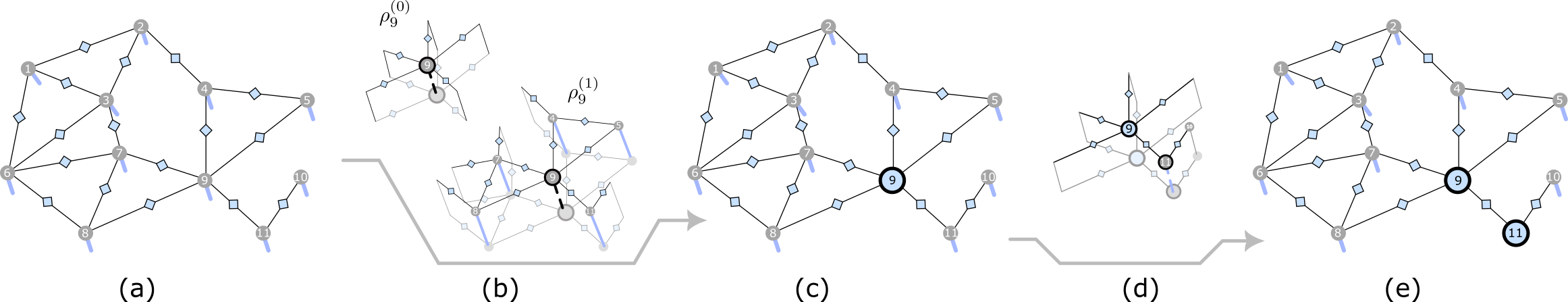}
    \caption{(a) A PEPS on some graph from which we must sample. (b) When sampling a particular site, e.g. site no. 9, we computed the reduced density matrix of this site by including a environment comprised of at most only first neighbors around the site. Shown are the tensor network expressions for the density matrices, $\rho_g^{(0)}$ and $\rho_g^{(1)}, $obtained without including any environment tensors and including only first neighbours, respectively. (c) We sample the spin no. 9 according to the chosen density matrix and  sliced the PEPS tensors at site 9 by fixing physical index to the sampled spin value, thus, reducing the PEPS tensor to a tensor (highlighted blue) without any physical index. (d)-(e) We then repeated the procedure on another site, here we chose site no. 11. The reduced density matrix of site no. 11 now depends on the updated tensor no. 9 obtained in the previous sampling step.}
    \label{fig:conditional_sampling}
\end{figure*}

In the MPS case, spins are conditionally sampled in a sweep starting from the left or right boundary of the MPS and then proceeding towards the other end.  At each step, we compute the reduced density matrix of the current site, and use the corresponding probability distribution to sample the local basis of that spin. We then fix the current physical index in the MPS tensor with the value of the drawn basis state, thus, reducing the MPS tensor to a matrix. We then multiply this matrix into the next tensor, and then iterate the above local sampling procedure. 

For a PEPS that is defined on an arbitrary geometry, as is the case in this paper, there's no canonical order (e.g. to left to right) in which to sample the sites. Therefore, in this case, we sample the spins in a random order. At each sampling step, we compute the reduced density matrix of the current spin. This density matrix may be computed using either the simple (= Identity), cluster, or full environment tensor, as described in the previous section. Fig. \ref{fig:conditional_sampling} illustrates the first two steps of the PEPS sampling procedure outlined above. In general, the accuracy of the reduced density matrix, and therefore of the sampling procedure, increases as progress from simple towards full environment, however, at increasing computational cost. In our benchmarking simulations, we chose a cluster of first-neighbour tensors and sampled 100 product states. The product state with the lowest energy was returned as the solution.

\section{Details of the PEPS simulations for QUBO problems}\label{app:simdetails}

\subsection{Classical Ising Hamiltonians (QUBO problems) studied in this paper} \label{app:problemset}

We considered two classes of classical Ising models for our benchmarking calculations: (1) randomly generated models and (2) models taken from the QPLIB database, which catalogs challenging optimization problems. From the QPLIB database we chose the problems listed in Table \ref{table:qplib}.
\begin{table}[ht]
\centering
\begin{tabular}{|c|c|c|}
\hline
Problem id & & \\
(this paper) & QPLIB identifier & Num. of spins  \\ \hline
Q1 & 3832 & 561 \\ \hline
Q2 & 3877 & 630 \\ \hline
Q3 & 3706 & 703 \\ \hline
Q4 & 3838 & 780 \\ \hline
Q5 & 3822 & 861 \\ \hline
Q6 & 3650 & 946 \\ \hline
Q7 & 3642 & 1035 \\ \hline
Q8 & 3693 & 1128 \\ \hline
Q9 & 3850 & 1225 \\ \hline
\end{tabular}
\caption{QPLIB problems studied in this paper.}
\label{table:qplib}
\end{table}

The Q-matrix of the QUBO formulation of all the classical Ising models used in this paper is available online on https://doi.org/10.5281/zenodo.12723409.

\subsection{PEPS parameters for the QUBO benchmarks}
Table \ref{table:pepsparams} lists the bond dimensions ($\chi$) and $\kappa$ values that we set in the flexible PEPS simulations of the QUBO benchmarks shown in Fig.\ref{fig:main_results}. We performed a bond dimension scaling for these problems, see Appendix \ref{app:bonddimscaling}; the bond dimension listed in Table \ref{table:pepsparams} corresponds to the scaling limit. For some problems setting $\kappa = 1$ was sufficient, which reduced the flexible PEPS optimization to an optimization over product states (equivalent to setting $\chi = 1$ in the fixed-PEPS setting). For some densely connected problems we obtained an accurate result already for $\kappa = 2$, in which case the flexible PEPS reduces to a flexible MPS.

\begin{table*}[t]
\centering
\begin{tabular}{|c|c|c|c|c|c|c|c|c|c||c|c|c|c|c|c|c|c|c|c|}
\hline
Method &  & R1 & R1 & R3 & R4 & R5 & R6 & R7 & R8 & Q1 & Q1 & Q3 & Q4 & Q5 & Q6 & Q7 & Q8 & Q9 \\ \hline
\multirow{2}{*}{Sim. Quantum Annealing} & $\kappa$ & 5 & 2 & 5 & 6 & 1 & 1 & 6 & 4 & 4 & 4 & 4 & 2 & 4 & 4 & 4 & 4 & 4\\ \cline{2-19} 
    & $\chi$ & 6 & 6 & 2 & 6 & 8 & 2 & 2 & 4 & 2 & 2 & 2 & 8 & 2 & 2 & 2 & 2 & 2\\ \hline \hline
\multirow{2}{*}{Hybrid Annealing} & $\kappa$ & 1 & 2 & 1 & 3 & 6 & 1 & 1 & 3 & 4 & 4 & 4 & 4 & 4 & 4 & 4 & 4 & 4\\ \cline{2-19}
     & $\chi$ & 2 & 2 & 2 & 4 & 2 & 2 & 2 & 2 & 4 & 2 & 6 & 4 & 4 & 6 & 2 & 2 & 4\\ \hline \hline
\multirow{2}{*}{Imaginary time evolution} & $\kappa$ & 3 & 2 & 1 & 6 & 1 & 3 & 4 & 4 & 4 & 4 & 3 & 4 & 4 & 4 & 3 & 4 & 4\\ \cline{2-19}
    & $\chi$ & 2 & 2 & 2 & 2 & 2 & 2 & 2 & 2 & 4 & 4 & 2 & 4 & 6 & 4 & 2 & 10 & 10\\ \hline
\end{tabular}
\caption{PEPS bond dimension ($\chi$) and $\kappa$ values used to obtain the results for the QUBO benchmarks shown Fig.\ref{fig:main_results}.}
\label{table:pepsparams}
\end{table*}

\subsection{Bond dimension scaling} \label{app:bonddimscaling}

\begin{figure}[t]
    \centering
    \includegraphics[width=\columnwidth]{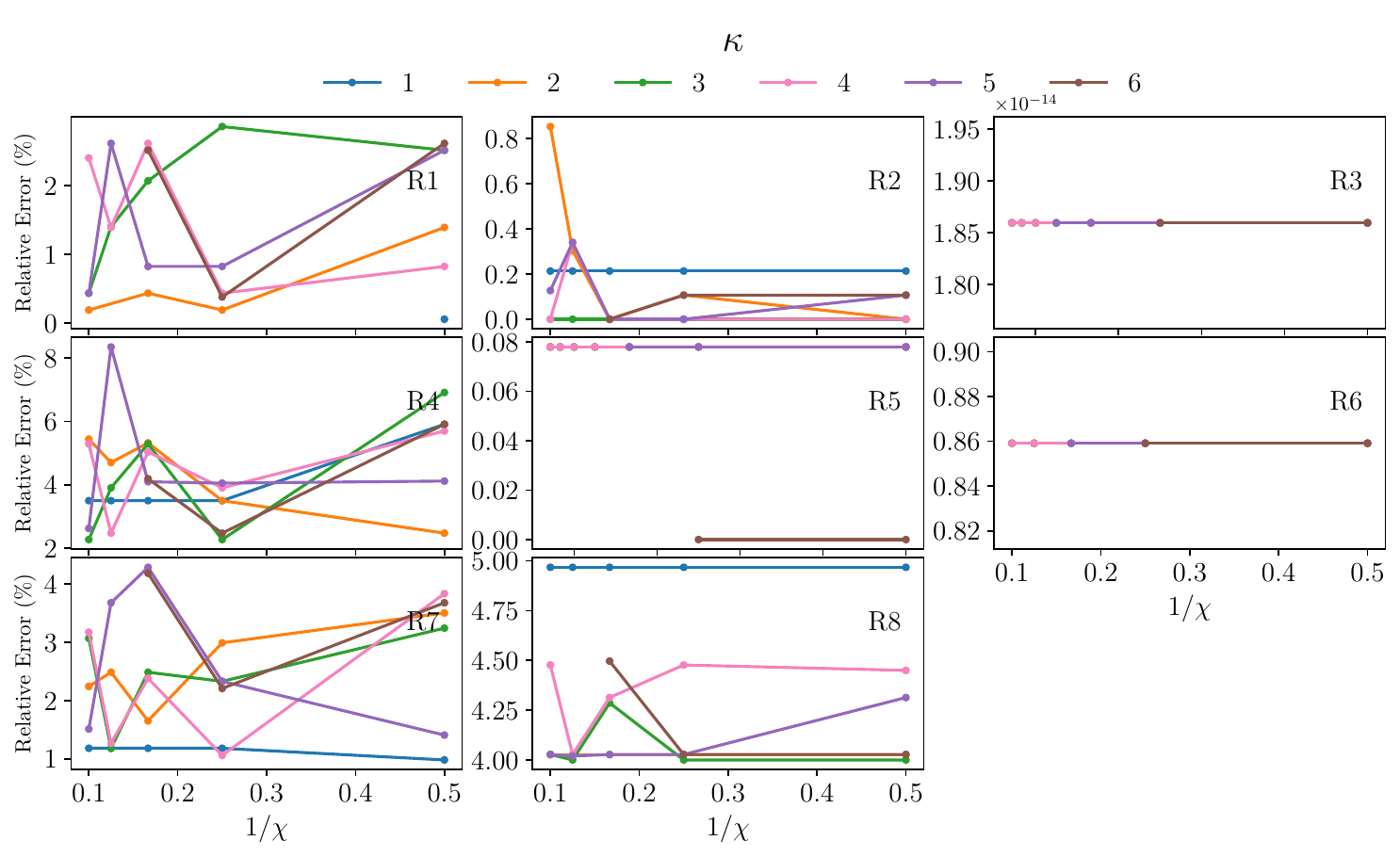}
    \caption{Bond dimension scaling for the random QUBO problems when using flexible PEPS hybrid annealing.}
    \label{fig:SQAh_Random}
\end{figure}

\begin{figure}[t]
    \centering
    \includegraphics[width=\columnwidth]{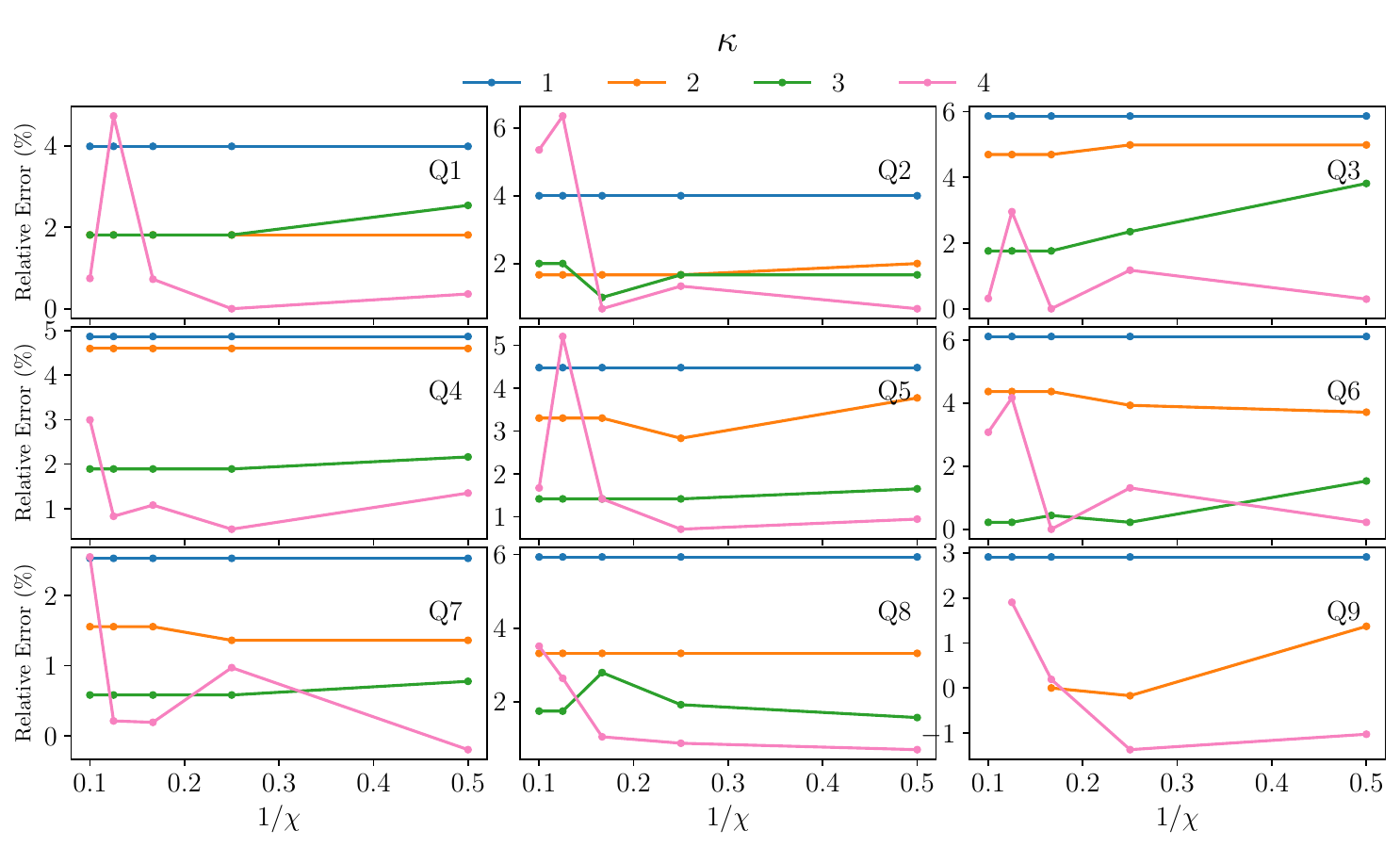}
    \caption{Bond dimension scaling for the QPLIB QUBO problems when using flexible PEPS hybrid annealing.}
    \label{fig:SQAh_QPLIB}
\end{figure}

\begin{figure}[t]
    \centering
    \includegraphics[width=\columnwidth]{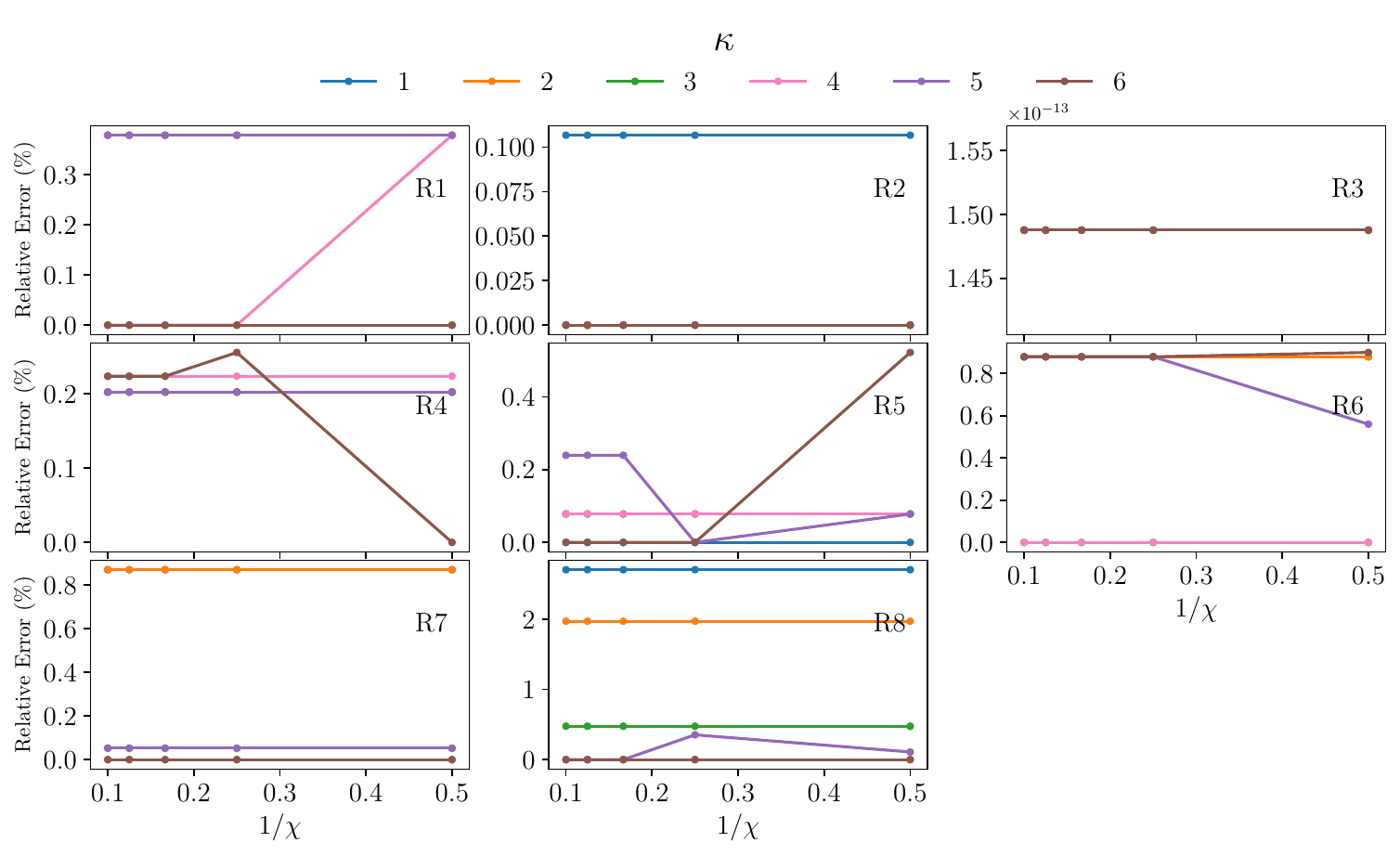}
    \caption{Bond dimension scaling for the random QUBO problems when using flexible PEPS imaginary time evolution.}
    \label{fig:Imag_Random}
\end{figure}

\begin{figure}[t]
    \centering
    \includegraphics[width=\columnwidth]{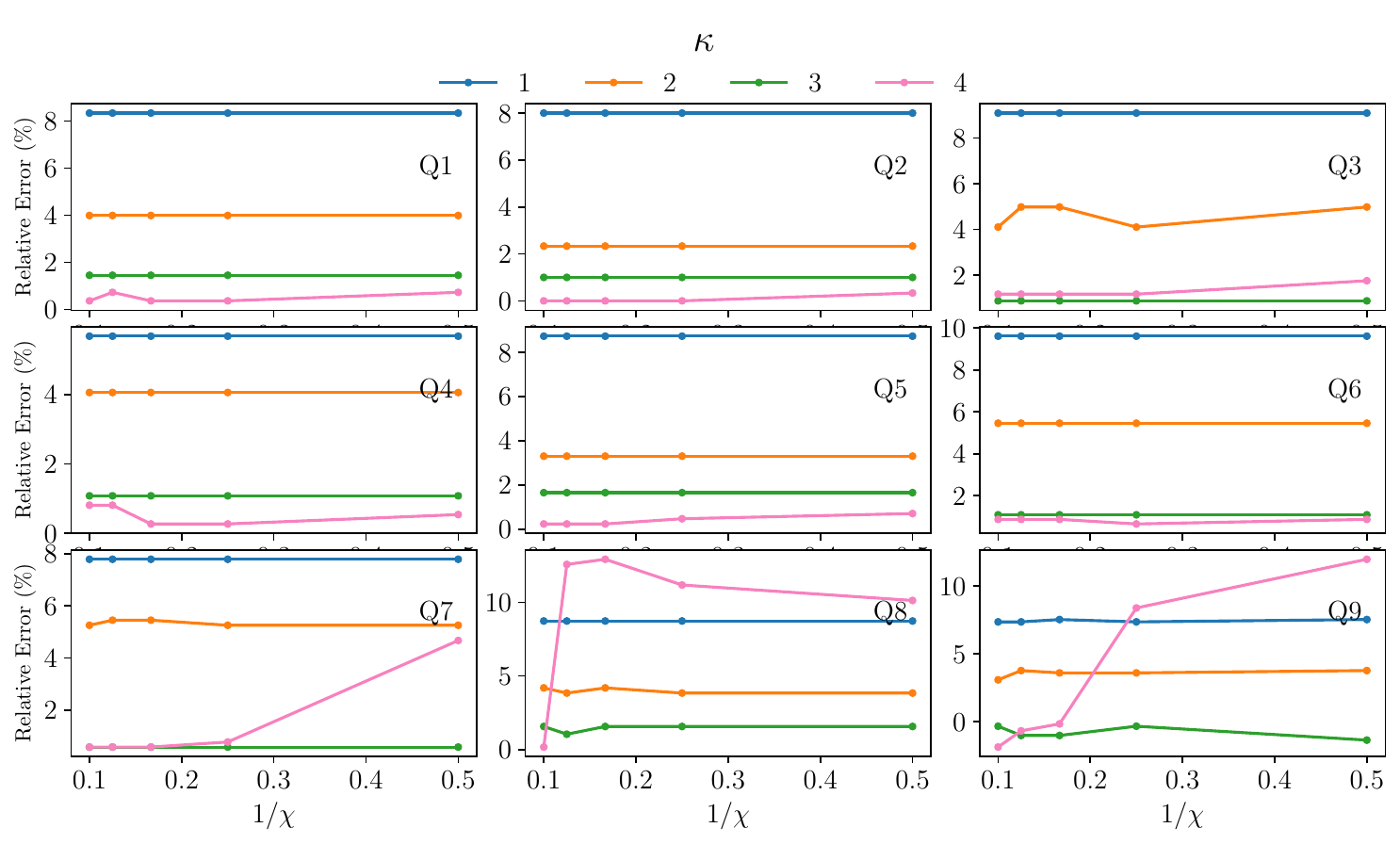}
    \caption{Bond dimension convergence for the QPLIB QUBO problems using flexible PEPS imaginary time evolution.}
    \label{fig:Imag_QPLIB}
\end{figure}

For the QUBO solutions reported in the main text using the flexible PEPS, we converged the result by increasing the bond dimension scaling for all the QUBO problems. Fig.~\ref{fig:SQAh_Random} and Fig.~\ref{fig:SQAh_QPLIB} shows the convergence of the solution with bond dimension for hybrid annealing for randomly generated and QPLIB QUBO problems respectively. Similarly, Fig.~\ref{fig:Imag_Random} and Fig.~\ref{fig:Imag_QPLIB} show the convergence with bond dimension for imaginary time evolution. Note the relatively high fluctuations in bond dimension scaling for higher $\kappa$ values. We believe these fluctuations are due to the limited accuracy of local measurements and could potentially be reduced by using cluster or full measurements.

\subsection{Annealing schedule} \label{app:annealingschedule}
\begin{figure}
    \centering
    \includegraphics[width=0.37\textwidth]{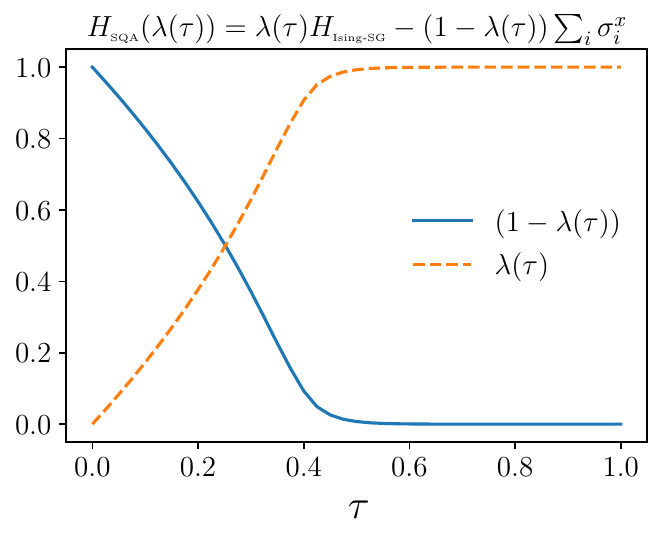}
    \caption{Schedule used for simulated quantum annealing.}
    \label{fig:sqa_schedule}
\end{figure}

In practice, the Hamiltonian should be deformed ``slowly enough'' relative to the time-scale equal to the inverse of the Hamiltonian gap. However, for large system, we do not know the gap in advance, and thus finding a suitable annealing schedule presents a key challenge to applying simulated annealing to find the ground state of the classical Ising Hamiltonian. Nonetheless, several effective heuristic annealing schedules have been proposed, such as the one shown in Fig.~\ref{fig:sqa_schedule}, which matches the schedule of the physical D-wave annealer. In our simulations, we simulated a discretized version of this schedule.

We assessed three different flexible-PEPS algorithms to solve the QUBO problems -- (1) Imaginary time evolution, (2) Simulated Quantum Annealing, and (3) Hybrid Annealing (these are described in the main text). We benchmarked these methods against the results obtained from Gurobi. The Gurobi time was limited to 300 seconds. The errors from the Gurobi solutions for the three methods (1) - (3) and the MIPGaps \footnote{The MIPGap returned by Gurobi estimates an upper bound on the error in the solution.} for the Gurobi solution for all problems are shown the Table~\ref{tab:error_values}.

\begin{table*}[ht]
    \centering
    \begin{tabular}{|c|c|c|c|c|}
        \hline
        \textbf{Prob. No.} & \textbf{Imag. time-evolution} & \textbf{Sim. Quantum Annealing} & \textbf{Hybrid Annealing} & \textbf{MIPGap Gurobi} \\
        \hline
        R1  & 0.00 & 69.54 & 0.06 & 0.07832 \\
        R2  & 0.00 & 70.24 & 0.00 & 0.04533 \\
        R3  & 0.00 & 21.79 & 0.00 & 0.08670 \\
        R4  & 0.00 & 74.15 & 2.27 & 0.07825 \\
        R5  & -0.00 & 1.73 & -0.00 & 0.10522 \\
        R6  & 0.00 & 22.63 & 0.86 & 0.12291 \\
        R7 & 0.00 & 68.28 & 0.98 & 0.09740 \\
        R8 & 0.00 & 68.51 & 4.00 & 0.00000 \\ \hline
        Q1 & 0.36 & 5.43 & 0.00 & 0.03623 \\
        Q2 & 0.00 & 5.00 & 0.67 & 0.02000 \\
        Q3 & 0.88 & 6.74 & 0.00 & 0.02053 \\
        Q4 & 0.27 & 5.95 & 0.54 & 0.03784 \\
        Q5 & 0.24 & 4.48 & 0.71 & 0.03302 \\
        Q6 & 0.66 & 4.15 & 0.00 & 0.03275 \\
        Q7 & 0.58 & 8.37 & -0.19 & 0.03502 \\
        Q8 & 0.17 & 5.24 & 0.70 & 0.03490 \\
        Q9 & -1.88 & 2.74 & -1.37 & 0.05651 \\
        \hline
    \end{tabular}
    \caption{Errors for  values for TEBD, SQA, SQAH, and the MIPGap for the Gurobi solution.}
    \label{tab:error_values}
\end{table*}

\subsection{Flexible PEPS convergence of simulated quantum annealing vs hybrid annealing} \label{app:convergence}

\begin{figure}
    \centering
    \includegraphics[width=\columnwidth]{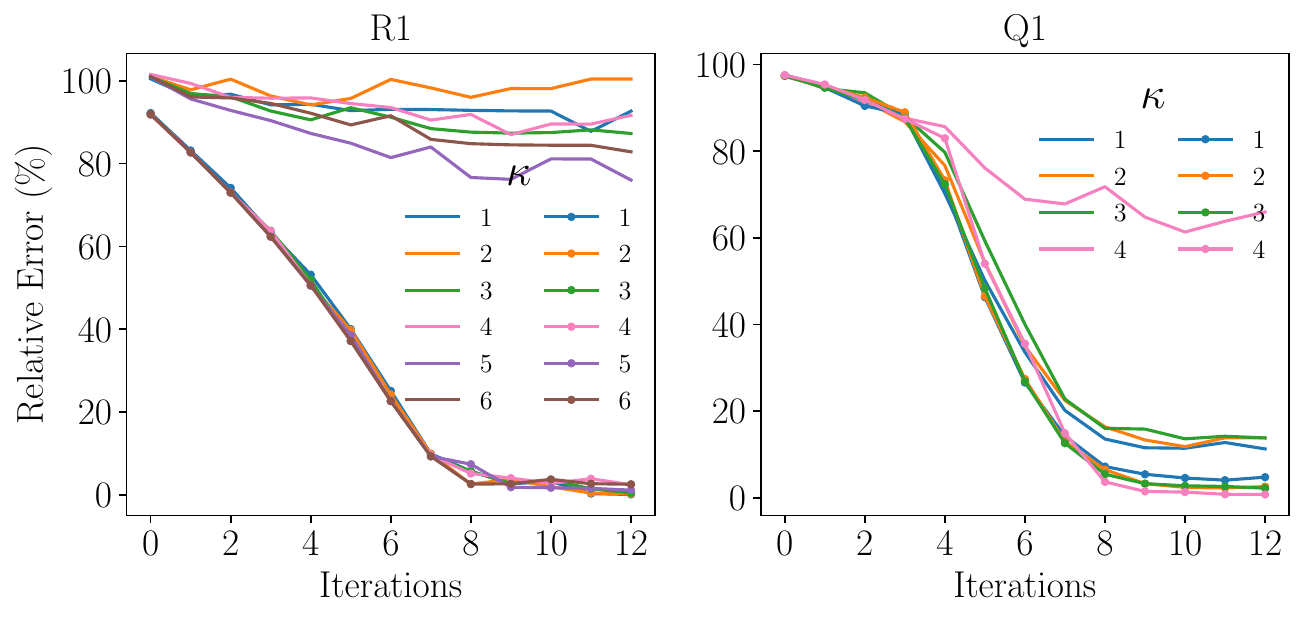}
    \caption{Comparison of the energy evolution of the simulated quantum and hybrid annealing for problems R1 and Q1. We see that Hybrid annealing produces a more accurate result and converges more smoothly.}
    \label{fig:comparison_energy_evolution_sqa_sqah}
\end{figure}

Simulated hybrid annealing usually performed better on the QUBO problems when compared to simulated quantum annealing. 
In Fig.~\ref{fig:comparison_energy_evolution_sqa_sqah}, we show the comparison more closely for two specific problems, the randomly generated QUBO R1 and the QPLIB problem Q1. Model R1 contains 80 spins on a densely connected graph with average vertex degree of $\approx 39$, while model Q1 is for a larger number of spins ($ = 561$ spins) but on a sparsely connected graph with average vertex degree $\approx 2$. We see that hybrid annealing is significantly more accurate than simulate quantum annealing, while also converging more smoothly.

\section{Critical PEPS geometries from the simulation of the uniform 2D quantum Ising model} \label{app:criticalPEPSgeometry}

\begin{figure}[b]
    \centering
    \includegraphics[width=\columnwidth]{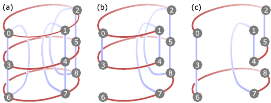}
    \caption{The critical PEPS geometries that emerge for different $\kappa$ values in the flexible PEPS ground state optimization of the 2D quantum transverse field Ising model on a square lattice: We found (a) a square lattice for $\kappa=4$, (b) an irregular graph for $\kappa=3$, and (c) a 1D chain with periodic boundaries for $\kappa=2$. The indicated order of the sites is the one we chose for the algorithm, the connectivity was dynamically determined by the flexible PEPS optimization.}
    \label{fig:uTFIM_graph}
\end{figure}

Fig.~\ref{fig:uTFIM_graph} shows an expanded view of the ``critical geometries'' shown in Fig.~\ref{fig:TFIM_critical_comparison}, which were obtained from flexible-PEPS simulation of uniform transverse-field Ising model on 2D \textit{square} lattice, with a $3\times 3$ unit cell with periodic boundaries. The flexible PEPS geometry changes along the optimization as edges are deleted whenever a tensor acquires more than $\kappa$ bond indices. The geometries shown are those acquired at the sharp transitions in $\langle \sigma_z (\lambda) \rangle$ seen in Fig.~\ref{fig:TFIM_critical_comparison} for $\kappa = 2,3,4$. For $\kappa = 4$, the sharp transition coincides with the critical point $\lambda \approx 3.1$ of the 2D quantum Ising model on a square lattice. At this point we indeed find that the flexible-PEPS acquires a square lattice geometry depicted in Fig.~\ref{fig:uTFIM_graph}(a). Note, however, the sites of the square lattice are ordered differently from how we have ordered them in the algorithm; the site-order emerges from the flexible PEPS optimization. For $\kappa = 2,3$ the flexible PEPS acquires the geometries shown in 
Fig.~\ref{fig:uTFIM_graph}(b), (c), respectively. As discussed in the main text, the sharp transitions seen in Fig.~\ref{fig:TFIM_critical_comparison} for these $\kappa$ values coincide with the critical points of uniform quantum Ising models defined on these geometries.  

\begin{figure}[t]
    \centering
    \includegraphics[width=\columnwidth]{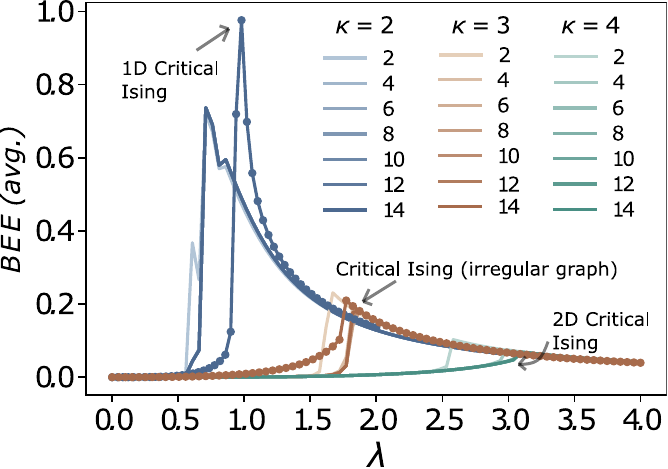}
    \caption{Average bond entanglement entropy vs. the strength of the magnetic field $\lambda$ for the ground state of the uniform 2D quantum transverse field Ising model $H_{\mbox{\tiny TFIM}}$ on a square lattice. Curves without markers correspond to \textit{flexible} PEPS simulations with $\kappa = 2,3,4$. For each $\kappa$, the results were converged for increasing bond dimension, lighter curves correspond to smaller bond dimensions, as listed in the legend for various $\kappa$. The curves with markers correspond to \textit{fixed} PEPS simulations of Ising models defined on the ``critical'' geometries shown in Fig.~\ref{fig:uTFIM_graph} that are recovered from the flexible-PEPS simulations.}
    \label{fig:TFIM_BEE_compare}
\end{figure}

In Fig.~\ref{fig:TFIM_BEE_compare}, we plot the average bond entanglement entropy (BEE; defined in the main text) in the ground state of $H_{\mbox{\tiny TFIM}}$ obtained using flexible PEPS for various values of the magnetic field $\lambda$. We find that BEE displays sharp transitions analogous to $\langle \sigma_Z \rangle$ in Fig.~\ref{fig:TFIM_critical_comparison}, though, the locations of the transitions are displaced for some curves, especially for the flexible PEPS curves for $\kappa =  2$. Nonetheless, these results illustrate that BEE, while being a simple heuristic metric, can be useful for locating critical points within a reasonable accuracy.  

\section{Flexible MPS}\label{app:flexiblemps}

\begin{figure}
    \centering
    \includegraphics[width=\columnwidth]{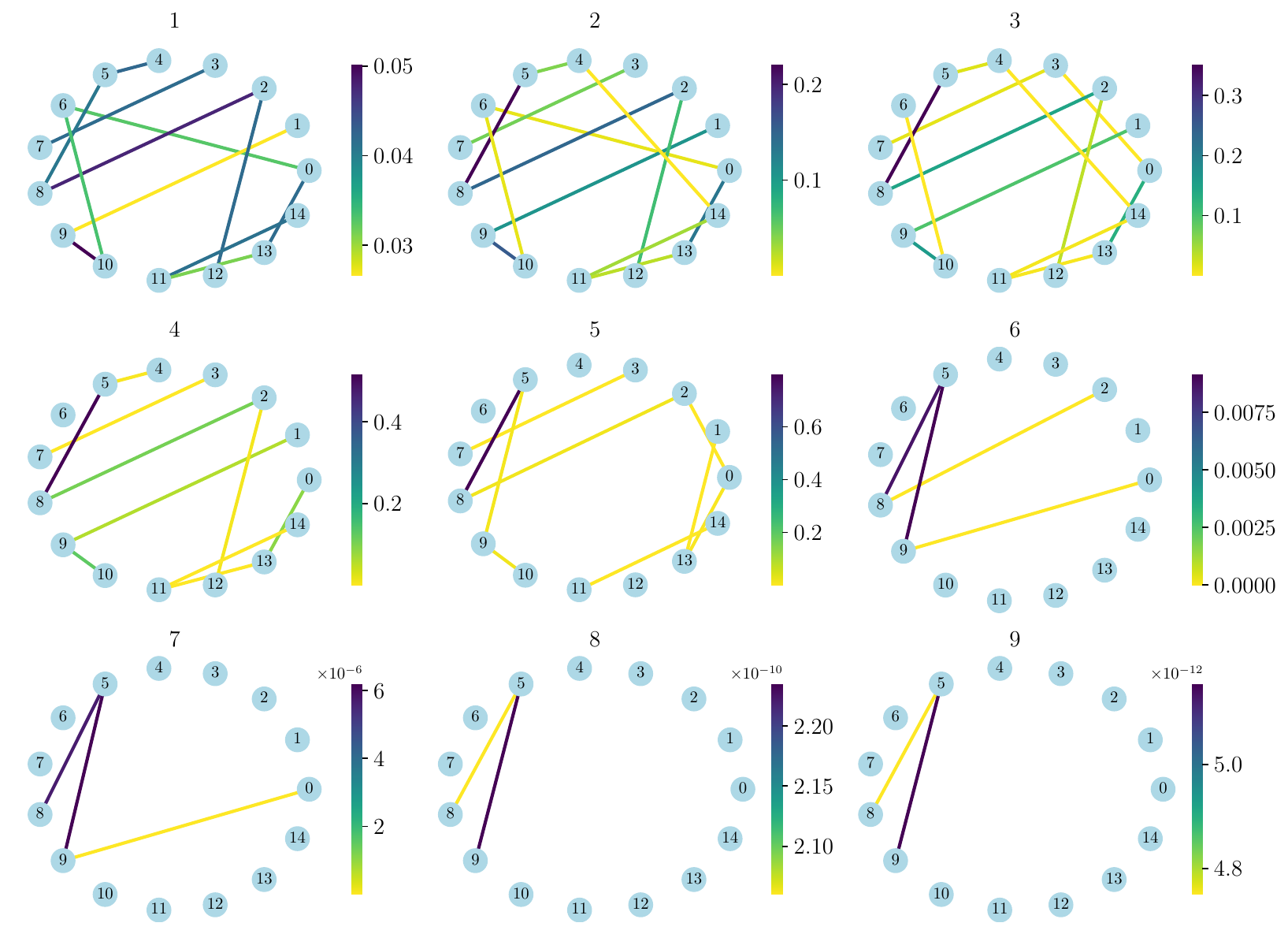}
    \caption{The imaginary-time evolution of the flexible PEPS geometry when optimizing for the ground state of a randomly generated classical Ising spin glass comprised of 15 spin, keeping $\kappa = 2$. Here we show the PEPS geometry at certain time steps selected during the evolution. (Physical indices of the PEPS are suppressed in this pictures.) We begin with the product state $\ket{+}$ (not shown in figure), which quickly becomes entangled after a few steps (panel 1). Applying more gates beegins to violate the vertex degree constraint, $\kappa = 2$, which triggers edge deletions. In this example, we found that the algorithm eventually converged to the ground product state.}
    \label{fig:graph_evolution}
\end{figure}

When $\kappa = 2$ (equivalent to at most two bond indices per tensor), the flexible PEPS algorithm described in this paper generates only MPSs (generally, a tensor product of several MPSs). Fig.~\ref{fig:graph_evolution} shows the evolution of the flexible PEPS geometry in imaginary time when optimizing for the ground state of a randomly generated classical Ising spin glass of 15 spins and keeping $\kappa = 2$. We found that after a few steps the optimization produces an MPS (Fig.~\ref{fig:graph_evolution}(i)) with an optimization-determined geometry, and that the bonds of this MPS gradually weaken, eventually converging to a product state.

\end{document}